\documentclass[a4paper,fleqn,usenatbib,referee]{mnras}
\usepackage{newtxtext,newtxmath}

\usepackage[T1]{fontenc}
\usepackage{ae,aecompl}

\usepackage{graphicx}	
\usepackage{amsmath}	
\usepackage{amssymb}	
\usepackage{commath}	

\usepackage{tikz}        
\usepackage{tikz-3dplot} 

\usepackage{bm}         

\newcommand{\rlight}{R_{\rm L}}
\newcommand{\ihat}{\textbf{\^\i}} 
\newcommand{\jhat}{\textbf{\^\j}}

\usepackage{float}      
\usepackage{booktabs}   
\usepackage{caption}    
\usepackage{subfig}     
\usepackage{epsfig}
\usepackage{xcolor}

\usepackage{tabularx}

\usepackage{epstopdf}   

\title[Pulsed emission from an off-centred magnetic dipole]{Pulsed emission from a rotating off-centred magnetic dipole in vacuum}
\author[Anu Kundu]{
Anu Kundu\thanks{E-mail: anu.kundu@astro.unistra.fr},
J\'er\^{o}me P\'etri
\\
Observatoire astronomique de Strasbourg, Universit\'e de Strasbourg, CNRS, UMR 7550,\\
11 rue de l'universit\'e, F-67000 Strasbourg, France
}
\date{Accepted XXX. Received YYY; in original form ZZZ}
\pubyear{2017}

\begin{document}
\label{firstpage}
\pagerange{\pageref{firstpage}--\pageref{lastpage}}
\maketitle

\begin{abstract}
The topology of the electromagnetic field around neutron stars severely impacts pulsar physics. While most of the works assume a standard centred dipolar magnetic field model, recently some efforts have been made to explain how inclusion of higher multipolar components could drastically change our understanding of these objects. Also, for simplicity, it has always been assumed that the magnetic moment coincides with the geometrical centre of the star. However, lately, a more general picture has been put forward in which the magnetic dipole moment is shifted off from the centre of the star. It has been demonstrated that the rotating off-centred dipole can be expanded into multipolar components. We study the effects of an off-centred rotating dipole on various characteristic emission features of pulsars in vacuum. The reliability of the off-centred case and its consequences on the magnetic field line structure, shape of the polar caps, high energy and radio emission phase plots and corresponding light curves along with a comparison with the standard centred case are discussed. It has been seen that an off-centred dipole breaks the north-south symmetry and allows for more flexibility in radio and high-energy light-curves fitting and phase lag.
\end{abstract}


\section{Introduction}
Pulsars are a special class of neutron stars \citep{gold_1968_gold_rotatingns_pulsars_1968} with strong magnetic fields typically up to an order of $10^{12}$~G and rotation period lying between 1.4~ms and 8.51 seconds. Since their discovery \citep{hewish_1967_hewish_pulsar_discovery_1968}, several attempts have been made to understand these mysterious objects. The magnetic field strength and topology of pulsars seems to play a central role in this investigation. They, indeed, provide insight into the physical processes like particle acceleration, radiation and emission mechanisms and hence, extensive literature focusses on it.

The electromagnetic field equations for a rotating dipole in vacuum were first presented by \cite{deutsch_1955_deutsch_the_1955} considering that the field is symmetrical about the magnetic axis which is inclined at an angle to the rotation axis of the star. One of the first pulsar models describing the magnetic poles and the magnetic field line structure was put forward by \cite{radhakrishnan_1969_radhakrishnan_polarization_1969} who compared the polarisation properties of a pulsar for different frequencies and proposed that the origin of emission must lie in the proximity of the poles. \cite{goldreich_1969_goldreich_pulsar_electrodynamics_1969} further discussed the electrodynamics of neutron stars for the simplest case which assumes magnetic dipole moment to be aligned with the rotation axis and assessed various properties of the region surrounding the star. Later, the inclined cases were considered and it was highlighted by \cite{michel_alignment_1970} and \cite{davis_magnetic-dipole_1970} that the radiation torque aligns the magnetic axis with the rotation axis. However, they assumed the star to be a perfectly conducting sphere. It was \cite{goldreich_1970_goldreich_mag_axis_alignment_1970} who investigated the alignment of a rotating magnetic dipole in a conducting, rigid, but non spherical body. \cite{pacini_1967_pacini_energy_1967} considered the oblique rotator model and discussed the energy emission from a neutron star and also presented the equations for the components of external electromagnetic field in vacuum \citep{pacini_1968_pacini_rotating_1968}. \cite{davis_1947_davis_stellar_1947} had discussed a possibility that the ion distribution in space modifies the electric and magnetic fields and hence, have a bearing on the electromagnetic field equations. He also suggested that both the fields extend far into the space and that the magnetic field cannot be purely dipolar.

Most of the literature which followed focused on these standard assumptions where a centred dipolar field was considered with its magnetic axis inclined at some angle with the rotation axis.

However, with the discovery of PSR J2144-3933 which has a period of 8.51 s, by far the longest of any known radio pulsar, the existing models were challenged and it was realised that this simple assumption needs revision \citep{young_radio_1999}. It was argued by \cite{gil_vacuum_2001} that to explain this extremal observation, a complicated magnetic field with multipolar components has to be considered. Soon, \cite{gil_modelling_2002} modelled the surface magnetic field of neutron stars which required strong and non-dipolar surface magnetic field near the pulsar polar cap. The electromagnetic field was considered to be a superposition of the global dipole field and a small scale magnetic anomaly. \cite{petri_multipolar_2015} included multipolar components to the electromagnetic field in a self-consistent way and demonstrated that working with only a dipole field can, indeed, be very misleading.

It is also assumed, by default, that the centre of the magnetic dipole coincides with the centre of the rotation axis i.e. with the centre of the pulsar. However, it has been shown by \cite{stift_1974_stift_the_1974} and \cite{komesaroff_1976_komesaroff_planet_multipole_1976} that a deviation from this centred assumption i.e. an off-centred geometry is possible for stars and planets respectively.

The offset idea was applied to neutron stars by \cite{harding_pulsar_2011} to investigate the effects of offset polar caps on pair cascades near the surface. Recently, \cite{petri_radiation_2016} studied the effect of an offset dipole anchored in the neutron star interior and calculated exact analytic solutions for the electromagnetic field in vacuum outside the star. We will be using these equations to study the consequences of the off-centred approach.

Some latest works have shown positive inclinations towards the multipolar field and the off-centred approach. \cite{archibald_high_2016} reported the observation of a pulsar with a high braking index. If it is not just a glitch and proven to be stable in future, one possible explanation for such high braking index would be the presence of higher order multipoles and could highlight the significance of considering higher components while studying pulsar spin down. \cite{barnard_effect_2016} modelled the pulsar light curves with an offset dipole prescription to fit them with the observations of Vela pulsar using different emission geometries. It was concluded that for small shifts($ \lesssim 20 \%$ of the radius of the pulsar) from the centre and constant emissivity, the offset polar cap dipole magnetic field is a good candidate and also favoured, but only to an extent, for larger shifts considering variable emissivity.

The aim of this paper is to study how consideration of an off-centred rotating dipole affects the emission features of pulsars in vacuum by outlining the comparison with the centred dipole and to discuss the major differences for an insight into the understanding of the emission process. The outline of the sections in the paper is as follows. We present an overview of the off-centred geometry and a description of the inclination angles and other parameters in Section~\ref{Off-centred geometry}. Section~\ref{Magnetic field lines structure} shows the structure of magnetic field lines for an orthogonal case. In Section~\ref{Results and discussions}, we discuss the results obtained for the shape of the polar caps (Section~\ref{Polar cap geometry}), make an analysis of the error associated with our approach in Section~\ref{Error analysis}, the radio and high energy emission phase plots (Section~\ref{Radio and high energy phase plots}) pursuing a goal similar to that of \cite{bai_uncertainties_2010} i.e. to study emission mechanism in vacuum. The phase plots are followed by the corresponding light curves in Section~\ref{Radio and high energy emission light curves} and in the end, we conclude all our results in Section~\ref{Conclusions}.

\section{Off-centred geometry}\label{Off-centred geometry}
In this section, we summarize the description of the characteristic parameters for an off-centred dipolar magnetic field and present its geometry.

The underlying assumption for an off-centred geometry is to consider the magnetic moment to be shifted away with respect to the geometrical centre of the star. While dealing with the oblique case for a centred geometry we need only one parameter denoted by $\chi$ which defines the obliquity of the star, but for an off-centred star we need to define three additional parameters for a complete geometrical description of the magnetic topology. The four characteristic parameters are defined as follows:
\begin{itemize}
 \item $\alpha$ represents the obliquity of the magnetic moment with respect to the rotation axis.
 \item $\beta$ is a measure of the projection of the magnetic moment out of the meridional plane defined by the plane $xOz$ where (O,x,y,z) represents the Cartesian coordinate system.
 \item $\delta$ is the angle between the rotation axis and the line joining the centre of star to the centre of the magnetic moment.
 \item $D$ is the distance between the stellar centre and the magnetic moment whose vector representation is $\mathbf{d} = D(\sin{\delta}, 0, \cos{\delta})$.
\end{itemize}
Two important parameters but not directly related to the pulsar are the inclination~$\zeta$ of the line of sight with respect to the rotation axis and the emission height~$h$ taking the stellar surface as a reference meaning zero altitude, $h=0$. The angle~$\zeta$ determines the light-curve profiles and whether radio and/or high energy photons are detected.

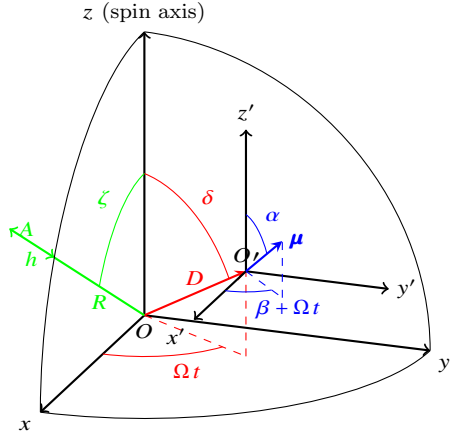
\begin{figure}
\centering


\tdplotsetmaincoords{70}{110}

\pgfmathsetmacro{\rvec}{.6}
\pgfmathsetmacro{\thetavec}{60}
\pgfmathsetmacro{\phivec}{60}

\begin{tikzpicture}[scale=4,tdplot_main_coords]

\coordinate (O) at (0,0,0);

\tdplotsetcoord{P}{\rvec}{\thetavec}{\phivec}

\tdplotsetcoord{A}{1}{60}{0}

\tdplotsetcoord{B}{1.5}{60}{0}


\draw[thick,->] (0,0,0) node[below] {$O$} -- (1,0,0) node[anchor=north east]{$x$};
\draw[thick,->] (0,0,0) -- (0,1,0) node[anchor=north west]{$y$};
\draw[thick,->] (0,0,0) -- (0,0,1) node[anchor=south]{$z$ (spin axis)};

\tdplotsetrotatedcoords{0}{0}{0}

\draw[-stealth,thick,color=red] (O) -- (P) node [midway, above] {$D$};
\draw[dashed,color=red,tdplot_rotated_coords] (0,0,0) -- (0.26,0.45,0);
\draw[dashed,color=red,tdplot_rotated_coords] (0.26,0.45,0) -- (0.26,0.45,.3);
\tdplotdrawarc[tdplot_rotated_coords,color=red]{(0,0,0)}{0.4}{0}{60}{anchor=north west}{$\Omega\,t$}

\draw[thick,color=green] (O) -- (A) node [midway, below] {$R$} ;
\draw[<->,thick,color=green] (A) -- (B) node [midway, below] {$h$} node [above, right] {$A$} ;



\tdplotsetthetaplanecoords{\phivec}

\tdplotdrawarc[red, tdplot_rotated_coords]{(0,0,0)}{0.5}{0}{\thetavec}{anchor=south west}{$\delta$}

\tdplotsetthetaplanecoords{0}

\tdplotdrawarc[green, tdplot_rotated_coords]{(0,0,0)}{0.5}{0}{60}{anchor=south east}{$\zeta$}


\tdplotsetrotatedcoords{0}{0}{0}

\tdplotsetrotatedcoordsorigin{(P)}

\draw[thick,tdplot_rotated_coords,->] (0,0,0) node[above] {$O\prime$} -- (.5,0,0) node[anchor=north east]{$x'$};
\draw[thick,tdplot_rotated_coords,->] (0,0,0) -- (0,.5,0) node[anchor=west]{$y'$};
\draw[thick,tdplot_rotated_coords,->] (0,0,0) -- (0,0,.5) node[anchor=south]{$z'$};


\draw[-stealth,thick,color=blue,tdplot_rotated_coords] (0,0,0) -- (.2,.2,.2) node [right] {$\pmb{\mu}$} ;
\draw[dashed,color=blue,tdplot_rotated_coords] (0,0,0) -- (.2,.2,0);
\draw[dashed,color=blue,tdplot_rotated_coords] (.2,.2,0) -- (.2,.2,.2);

\tdplotdrawarc[tdplot_rotated_coords,color=blue]{(0,0,0)}{0.2}{0}{45}{anchor=north west}{$\beta+\Omega\,t$}

\tdplotsetrotatedthetaplanecoords{45}

\tdplotdrawarc[tdplot_rotated_coords,color=blue]{(0,0,0)}{0.2}{0}{55}{anchor=south west}{$\alpha$}

\begin{scope}[canvas is xy plane at z=0]
     \draw (1,0) arc (0:90:1);
\end{scope}
\begin{scope}[canvas is xz plane at y=0]
     \draw (1,0) arc (0:90:1);
\end{scope}
\begin{scope}[canvas is yz plane at x=0]
     \draw (1,0) arc (0:90:1);
\end{scope}
  
\end{tikzpicture}
\caption{Geometry of the off-centred dipole showing the three important angles $\{\alpha, \beta, \delta\}$ and the distance $D$. Two additional parameters related to observations are the line of sight inclination~$\zeta$ and the emission height~$h$. The plot corresponds to time~$t$ assuming that $\bmu$ lies in the $(xOz)$ plane at $t=0$. The figure is adopted from \protect\cite{petri_polarized_2017-1}.}
\label{fig:off_star_geometry}
\end{figure}

Fig.~\ref{fig:off_star_geometry} (adopted from \cite{petri_polarized_2017-1}) depicts the geometry of an off-centred star of radius $R$. Because the electromagnetic field equations under consideration are with respect to the rotation axis of the pulsar, we need to make our calculations in the same frame. However, all our information about the magnetic moment and inclinations is in the magnetic axis frame. Therefore, we always begin from the latter frame and make a transformation to the former to proceed with the calculations. Hence, two reference frames are considered which are described as follows.

First is the unprimed reference frame $xyz$ whose centre coincides with the geometrical centre $O$ of the star, which means it is also the frame attached to the rotation (spin) axis of star. The position vector for any point located inside the sphere of radius $R$ in this centred frame is defined as $\mathbf{r} = r\mathbf{n}$ with $ \mathbf{n} = (\sin{\theta}\cos{\phi}, \sin{\theta}\sin{\phi}, \cos{\theta}) $ where $r$ is the spherical radius and $ \theta $ and $ \phi $ are the polar and azimuth angles respectively.

Second is the primed reference frame $x^{\prime}y^{\prime}z^{\prime}$ whose centre coincides with the magnetic moment of the star which is located at $O^{\prime}$. The coordinate axis $x, y$ and $z$ remain parallel to the coordinate axis $x^{\prime}, y^{\prime}$ and $z^{\prime}$ respectively. This primed reference frame $x^{\prime}y^{\prime}z^{\prime}$ will be referred to as off-centred frame now on wards. The magnetic moment $ \pmb{\mu} = \mu \mathbf{m} $ ($\mathbf{m}$ is a unit vector) is shown in blue with polar angle and azimuth angle as $\alpha$ and $\beta$ respectively as seen from the off-centred frame. It is located at $\mathbf{d}$ and directed along the unit vector $ \mathbf{m} = (\sin{\alpha}\cos{\beta}, \sin{\alpha}\sin{\beta}, \cos{\alpha}) $. Our calculations correspond to time~$t=0$ and hence, instead of $\beta+\Omega\,t$, where $\Omega$ is the angular velocity, we always use $\beta$ for phase.

The distance between the origins of two frames, $D$ and other characteristic parameters are also shown in the figure. An important condition used throughout is $D<R$ as the magnetic moment must remain within the star. We define a normalised quantity $\epsilon = D/R$ to keep this in consideration and hence, we have, $\epsilon \in [0, 1[$. We do not include the point $\epsilon=1$ because in that case, at the surface, the magnetic field strength would diverge.

The lowest order correction to the dipole corresponds to a quadrupolar moment which is combined to the centred dipole in order to get an accurate analytical description of the electromagnetic field outside the star when $D\ll R$. In our work, we neglect higher order multipolar perturbations but in principle they can be added if higher precision is required, especially when $D\lesssim R$. The exact analytical expressions for dipolar and quadrupolar electromagnetic fields recently published in Appendix C of \cite{petri_radiation_2016} are used for all the calculations. We have implemented these exact analytical expressions using the spherical Hankel functions~$h_\ell^{(1)}(x)$. Thus, our code can handle pulsars with any period from millisecond to second, our solution generalising the Deutsch field without restriction on the ratio $R/\rlight$ ($\rlight$ is the radius of the light cylinder defined by last closed magnetic field lines and equals $c/\Omega$, where $c$ is the speed of light).

The angles $\alpha, \beta$ and $\delta$ are shuffled to get different orientations for the off-centred geometry. As the parameter space defined by the off-centred dipole is at least five dimensional, it is impractical to explore the full range of inclination angles and distances. We restrict our study to a sample of judiciously chosen configurations to emphasise the discrepancies between centred and off-centred geometry. Throughout the paper, the pulsar period is set to $R/\rlight=0.1$ for every topology, corresponding approximately to a 2~ms pulsar, and $\epsilon = 0.2$ is considered for the off-centred calculations. The former is chosen so because lower the ratio $R/\rlight$, higher is the resolution (which depends on the radius of the star) which increases the simulation time significantly. The latter is chosen to have a large shift from the centre but not too large as the error scales with increase in $\epsilon$ as $\epsilon^2$ (see Table~\ref{tab:error_analysis}).

\section{Magnetic field lines structure}\label{Magnetic field lines structure}
Our work investigates the implication of the magnetic topology on several important characteristics of the multi-wavelength pulsed emission. A good intuitive picture of the magnetic field lines will help to interpret the results discussed in subsequent sections. So, in this section we focus on the structure of the magnetic field lines. We also show the structure of the individual components of the magnetic field to visualize their independent contributions to the total electromagnetic field.

As specified in Section~$3.2$ of \cite{petri_radiation_2016}, the expressions for the magnetic and the electric field have been decomposed into the dipolar and the quadrupolar components both normalised separately and independently with some well defined chosen weights for each component. Note that this normalisation is arbitrary but must remain consistent with the coefficients of the expansions presented below. The expression for the field is as shown below:

\begin{equation}
\begin{split}
\label{eq:Foff}
\textbf{F}^{\rm off} = \textbf{F}^{\rm dip} (\psi \rightarrow \psi - \beta) + \epsilon [ (2 \cos{\alpha} \cos{\delta} - \sin{\alpha} \sin{\delta} \cos{\beta}) \textbf{F}^{\rm quad}_{m=0}(\psi &)\\
+ \cos{\alpha} \sin{\delta} \textbf{F}^{\rm quad}_{m=1}(\psi) + \sin{\alpha} \cos{\delta} \textbf{F}^{\rm quad}_{m=1}(\psi \rightarrow \psi - \beta &) \\
+ \sin{\alpha}\sin{\delta} \textbf{F}^{\rm quad}_{m=2}(2\psi \rightarrow 2\psi - \beta ) & ].
\end{split}
\end{equation}
$\textbf{F}$ represents the field (electric displacement $\textbf{D}$ or magnetic field $\textbf{B}$) and $\textbf{F}^{\rm dip}$ and $\textbf{F}^{\rm quad}$ represents the dipolar and the quadrupolar components of the field respectively. $\alpha$, $\beta$ and $\delta$ are the angles used to determine the geometry of the system as described in the previous section. The subscript $m$ represents the azimuthal mode to be considered. $\psi=\phi-\Omega\,t$ represents the actual phase (at time $t$). The change $( \psi \rightarrow \psi - \beta) $ corresponds to a simple shift in the azimuthal direction as expected from the orientation of the magnetic moment with respect to the $\textit{xOz}$ plane.

\begin{figure}
  \centering
   \resizebox{0.95\textwidth}{!}{\input{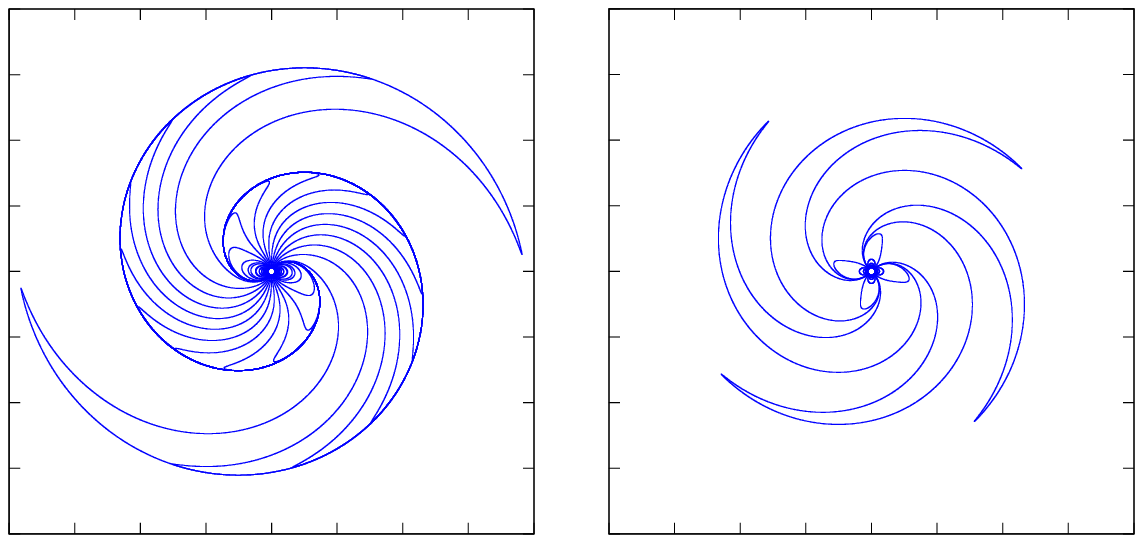}}
   \caption{The individual contribution of the dipolar (left) and the quadrupolar (right) components to the magnetic field line structure for an orthogonal off-centred case with $(\alpha, \beta, \delta) = (90\degr, 90\degr, 90\degr)$ and $\epsilon=0.2$.}
     \label{fig:mfleq_dip_quad}
\end{figure}

Displaying the full 3D geometry of magnetic field line is hard to lay on a 2D paper sheet. Thus, to simplify things for visualisation purposes, we present only the equatorial magnetic field lines structure. Indeed for special topologies, some magnetic field lines reside fully in the equatorial plane from the stellar surface to large distance. This can be achieved by setting the angles to appropriate values. A good sample is given by the subsequent plots showing the orthogonal geometry for two cases; $(\alpha, \beta, \delta) = (90\degr, 0\degr, 90\degr), (90\degr, 90\degr, 90\degr$). An example of individual contributions of both the dipolar and the quadrupolar components for an orthogonal off-centred case where $\beta=90\degr$ and $\epsilon=0.2$ is shown in Fig.~\ref{fig:mfleq_dip_quad}. We distinguish the basic property of a two-armed and a four-armed spiral pattern with quantum numbers respectively $(\ell,m)=(1,1)$ and $(\ell,m)=(2,2)$.

\begin{figure}
  \centering
   \resizebox{0.95\textwidth}{!}{\input{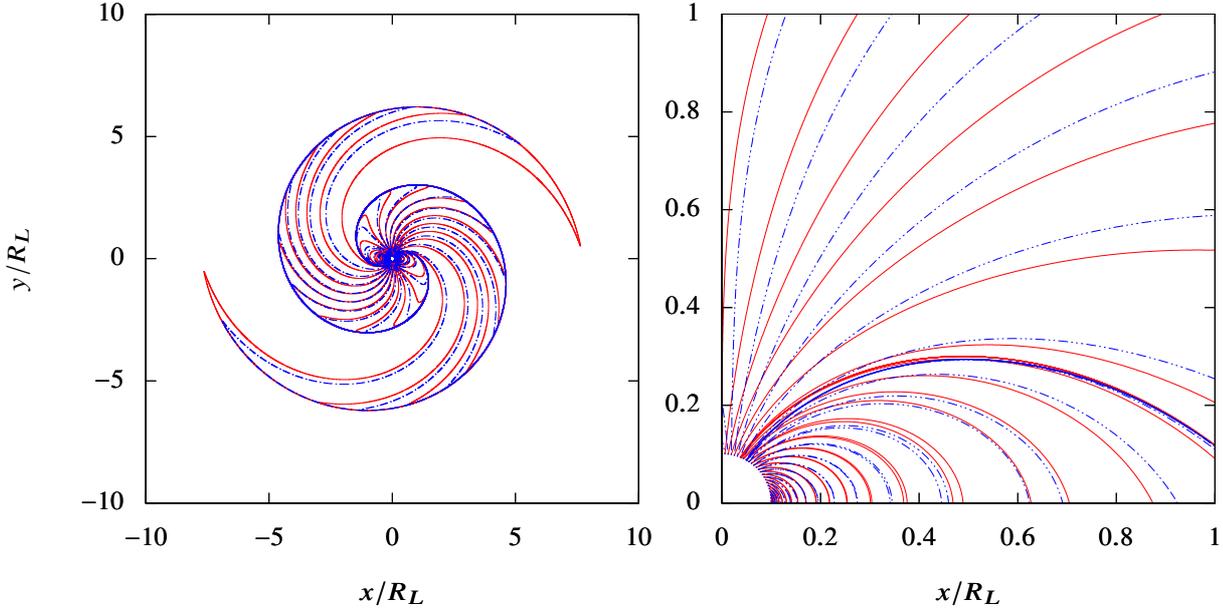}}
    \caption{Magnetic field line structure in the equatorial plane for orthogonal ($(\alpha, \beta, \delta) = (90\degr, 90\degr, 90\degr)$) centred case ($\epsilon=0$, in red) and off-centred case ($\epsilon=0.2$, in blue).}
      \label{fig:mfleq_beta_90}
\end{figure}

When both components, dipolar and quadrupolar, are added with appropriate weights according to Eqn.~(\ref{eq:Foff}), we get a faithful representation of a radiating off-centred dipole as long as the shift remains small $\epsilon\ll1$. The resulting structure of the magnetic field lines for the equatorial case where $\beta = 90\degr $ are compared for the two cases $\epsilon=0$ and $\epsilon=0.2$ as shown in Fig. \ref{fig:mfleq_beta_90} in red and blue respectively. Comparing the field pattern for the two cases, it is found that the spiral arms corresponding to both the poles are symmetric for the centred case, while they show an asymmetry for the other one. Also, a comparable shift in the entire field line geometry for both cases is prominent. To highlight this difference, we zoom in to a small section of the first quadrant shown on the left where the shift between the two cases is more clearly visible.

\begin{figure}
  \centering
   \resizebox{0.95\textwidth}{!}{\input{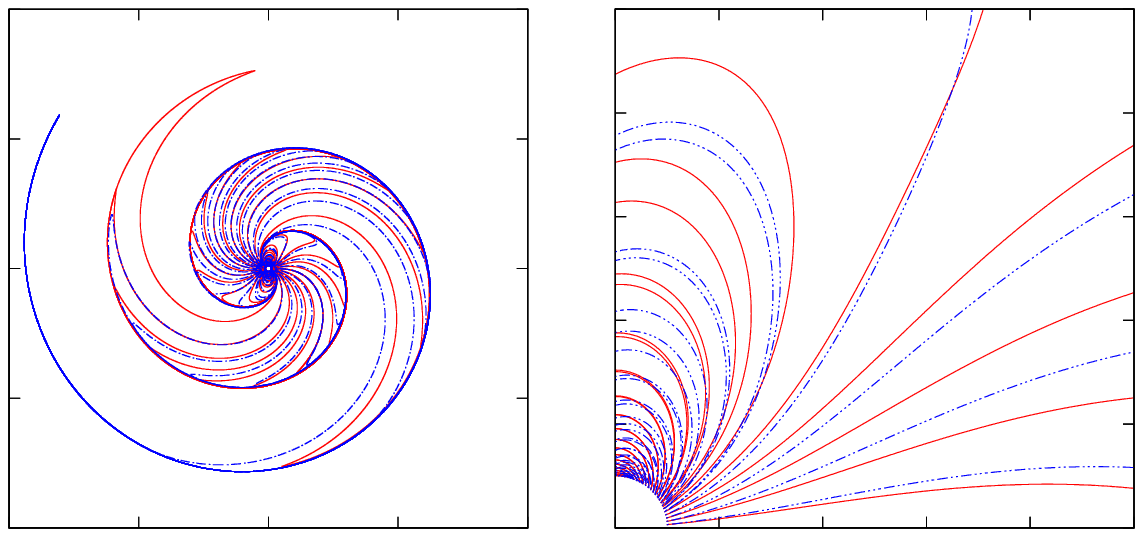}}
    \caption{Magnetic field line structure in the equatorial plane for orthogonal ($(\alpha, \beta, \delta) = (90\degr, 0\degr, 90\degr)$) centred case ($\epsilon=0$, in red) and off-centred ($\epsilon=0.2$, in blue).}
      \label{fig:mfleq_beta_0}
\end{figure}

The same is checked for another orthogonal configuration with $\beta = 0\degr $ in Fig.~\ref{fig:mfleq_beta_0}. We realise that the spiral symmetry which was present in centred case but missing in off-centred case for $\beta = 90\degr$, follows the same criteria here. Although, the asymmetry for the off-centred case (shown in blue) is much more pronounced than that in the previous case. We also see the shift mentioned for $\beta = 90\degr$ persisting here at large distances. As already stated in \cite{petri_radiation_2016}, this asymmetry would reflect even in the region outside the light cylinder through the striped wind originating from inside. The right side, again, zooms in to a better resolution to visualize the shift.

\section{Results and discussions}\label{Results and discussions}
Studying magnetic field lines is only glancing at the tip of the iceberg. Major observational consequences of our modified topology are thoroughly investigated in this section. Shape of the polar caps, considered to be the locus of the feet of the last closed field lines, is central to the problem of the pulsed emission. Pulsed emission is expected from sites of efficient particle acceleration where the electric field component parallel to the magnetic field is significant. These regions are suspected to be closely related to the locus of last closed field lines. Several attempts to localize these sites have been made since the discovery of the first pulsar in 1967. The question of emission close to the surface and/or close to the light-cylinder was debated until the launch of Fermi telescope in June 2008. Spectral analysis showed that high-energy emission must be located closer to the light-cylinder \citep{the_fermi-lat_collaboration_second_2013} in order to observe a sub-exponential cut-off. Moreover, the misaligned radio and gamma-ray pulses gives an indication about separated emission sites. Thus radio photons are probably produced closer to the surface than are gamma-ray photons. Indeed, the polar cap model \citep{sturrock_model_1971} suggests that the radio emission emanates mostly from the field lines on the stellar surface enclosing the polar caps with possible core and conal emission \citep{backer_pulsar_1976, rankin_toward_1983}. Farther away, slot gap models have been designed \citep{arons_pair_1983-1} to explain the high energy emission, photons are assumed to be produced in the vicinity of the last closed field lines from the surface up to the light cylinder. Some other variants also exist like an extended volume depicted by the outer gaps \citep{cheng_energetic_1986} and the annular gaps \citep{qiao_inner_2004}. We do not consider them further in this paper. The differences between radio and high-energy light-curves are now pointed out with several predictions based on phase plots, light-curves and phase lag between radio and high-energy peaks.

\subsection{Polar cap location and geometry} \label{Polar cap geometry}
The first step in understanding the polar cap geometry requires the knowledge of the location of the magnetic poles i.e. points of the intersection of the magnetic axis with the surface of the pulsar. To determine the coordinates of the poles, we employ the following method. The distance from the magnetic moment to the centre of the pulsar is $D$, as shown in Fig.~\ref{fig:off_star_geometry}. It is represented by the vector
\begin{equation}\label{eq:d_def}
 \textbf{d} = D(\sin{\delta}, 0, \cos{\delta}).
\end{equation}
The direction of the magnetic moment axis is 
\begin{equation}\label{eq:mm_def}
 \textbf{m} = (\sin{\alpha}\cos{\beta}, \sin{\alpha}\sin{\beta}, \cos{\alpha}).  
\end{equation}
Using Eqn.~(\ref{eq:d_def}) and Eqn.~(\ref{eq:mm_def}) and the general equation for a straight line, we can write the equation for the magnetic axis as 
\begin{equation}\label{eq:mag_axis}
\textbf{r} = \textbf{d} + \lambda \textbf{m}
\end{equation} 
where $\lambda$ is a parameter defined to find coordinates. Considering Cartesian coordinates with $\ihat$, $\jhat$, $\hat{\textbf{k}}$ being the unit vectors for $x$, $y$ and $z$ direction respectively, the position vector is written as $\textbf{r} = x \ihat + y \jhat + z \hat{\textbf{k}}$. Comparing this with Eqn.~(\ref{eq:mag_axis}), we get
\begin{equation} \label{eq:mag_axis_coordinates}
 x = D \sin{\delta} + \lambda \sin{\alpha} \cos{\beta}\\
 y = \lambda \sin{\alpha} \sin{\beta}\\
 z = D \cos{\delta} + \lambda \cos{\alpha}.
\end{equation}
To find the intersection of the magnetic axis and pulsar we substitute above coordinates representing all the points lying on a magnetic axis in the equation of a sphere of radius $R$ i.e. in $x^2 + y^2 + z^2 - R^2 = 0$;
\begin{align*}
& \Rightarrow(D \sin{\delta} + \lambda \sin{\alpha} \cos{\beta})^2 + (\lambda \sin{\alpha} \sin{\beta})^2 + (D \cos{\delta} + \lambda \cos{\alpha})^2 - R^2 = 0\\
& \Rightarrow \lambda^2 + 2D(\cos{\alpha}\cos{\delta} + \sin{\alpha}\sin{\delta}\cos{\beta})\lambda + (D^2 - R^2) = 0.
\end{align*}
The equation above represents a quadratic equation of form $ a\lambda^2 + b \lambda + c = 0 $ in $\lambda$ with\\
$ a = 1 $,\\
$b = 2D(\cos{\alpha}\cos{\delta} + \sin{\alpha}\sin{\delta}\cos{\beta})$,\\
$c = D^2 - R^2 $.\\
Considering only real roots for the above quadratic equation in $\lambda$ i.e. roots satisfying $ b^2 - 4ac \geq 0 $ which is equivalent to the condition $D < R$, we get:
\begin{equation}\label{eq:lambda}
 \lambda = \frac{( -b \pm \sqrt{b^2 - 4ac} )}{2a}.
\end{equation}
Using Eqn.~(\ref{eq:lambda}) in Eqn.~(\ref{eq:mag_axis_coordinates}), we get coordinates for the two poles.

\begin{figure}
  \centering
   \begin{tabular}{c}
   \includegraphics[width=0.95\textwidth]{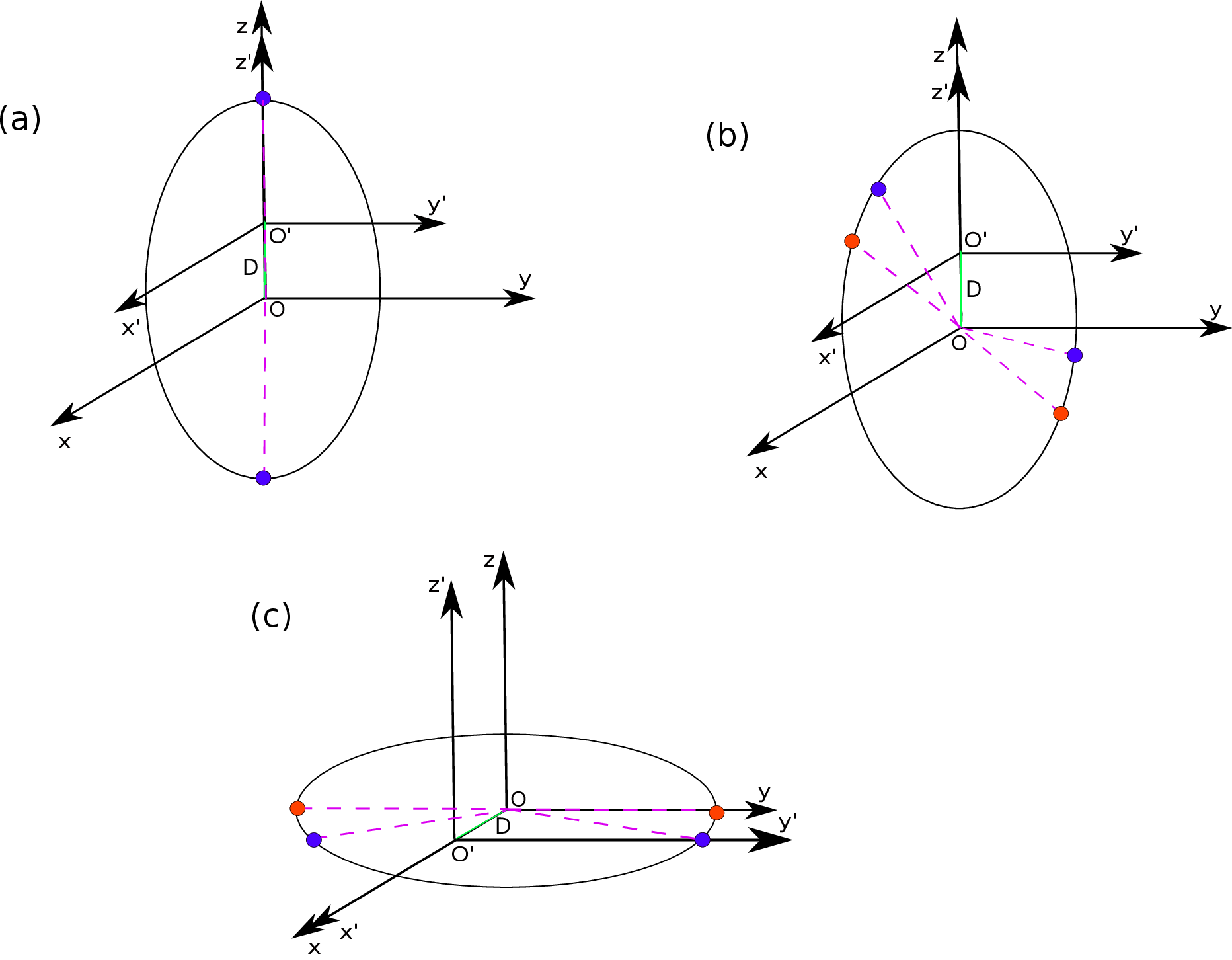}
  \end{tabular}
  \caption{A geometrical description for the location of the poles for various geometries; $(\alpha, \beta, \delta) = (0\degr, 0\degr, 0\degr)$ in (a), $(30\degr, 0\degr, 0\degr)$ in (b) and $(90\degr, 90\degr, 90\degr)$ in (c). The $xyz$ and the $x^{\prime}y^{\prime}z^{\prime}$ set of coordinate axis represent the centred and the off-centred frame with their centres being $O$ and $O^{\prime}$ respectively. The green solid line shows the shift $D$ between the two geometries. The black ring is a section of the surface of the pulsar in $xz$ plane for (a) and (b) while in $xy$ plane for (c). The solid-fill dots are the pole positions for centred (in red) and off-centred cases (in blue). Dashed magenta lines highlight the angles for the poles with respect to the rotation axis frame. Figures not to scale and provide only a rough depiction.}
  \label{fig:pc_geom}
\end{figure}

Following this procedure, we calculate the location of the poles and present them in the Table~\ref{tab:pole_location} for three particular cases for both; the centred and the off-centred geometries. It is interesting to note that the difference in azimuth of the two poles for $(\alpha, \beta, \delta) = (90\degr, 90\degr, 90\degr)$ gives a phase difference of exactly $180\degr$ for the centred case ($\epsilon = 0$) while it is higher than that for the off-centred case ($\epsilon = 0.2$). This difference is affiliated to the geometry. We make an attempt in Fig.~\ref{fig:pc_geom} to geometrically visualise these locations.

\begin{table}
\centering
\begin{center}
\begin{tabular}{l l l l l l}
\toprule
\boldmath{($\alpha, \beta, \delta$)} & \textbf{Pole 1} & & \textbf{Pole 2} & &  \boldmath{$\bigtriangleup \phi$}\\
 & \boldmath{$\theta_{1}$} & \boldmath{$\phi_{1}$} & \boldmath{$\theta_{2}$} & \boldmath{$\phi_{2}$} & \boldmath{$\phi_{2} - \phi_{1}$} \\
\midrule
 & \boldmath{$\epsilon=0$}\\
($0\degr,0\degr,0\degr$) & $0\degr$ & $0\degr$ & $180\degr$ & $0\degr$ & $0\degr$\\
($30\degr,0\degr,0\degr$) & $30\degr$ & $0\degr$ & $150\degr$ & $180\degr$ & $180\degr$\\
($90\degr,90\degr,90\degr$) & $90\degr$ & $90\degr$ & $90\degr$ & $270\degr$ & $180\degr$\\
\midrule
 & \boldmath{$\epsilon=0.2$}\\
($0\degr,0\degr,0\degr$) & $0\degr$ & $0\degr$ & $180\degr$ & $0\degr$ & $0\degr$\\
($30\degr,0\degr,0\degr$) & $24\degr$ & $0\degr$ & $144\degr$ & $180\degr$ & $180\degr$\\
($90\degr,90\degr,90\degr$) & $90\degr$ & $78\degr$ & $90\degr$ & $282\degr$ & $203\degr$\\
\bottomrule
\end{tabular}
\end{center}
\caption{Location of poles in terms of polar angle $\theta$ and azimuth angle $\phi$ (both rounded off), for various configurations defined by $(\alpha, \beta, \delta)$.}
\label{tab:pole_location}
\end{table}

In the figure, we show three cases with $(\alpha, \beta, \delta) = (0\degr, 0\degr, 0\degr)$ in (a), $(30\degr, 0\degr, 0\degr)$ in (b) and $(90\degr, 90\degr, 90\degr)$ in (c). The $xyz$ and the $x^{\prime}y^{\prime}z^{\prime}$ set of coordinate axis represent the centred and the off-centred frame with their centres being $O$ and $O^{\prime}$ respectively. Shift $D$ between two frames is shown in green. The black ring is a section of the surface of the pulsar in $xz$ plane for (a) and (b) while in $xy$ plane for (c) chosen based on location of the poles. The colour scheme is the same as used throughout the paper with red representing the centred case while blue representing the off-centred case. The solid-fill dots are the location of the poles. The magenta dashed lines highlight the angles for the poles with respect to the unprimed i.e. the rotation axis frame and are plotted based on the values given in the Table~\ref{tab:pole_location}. The representation is not to scale and represent only a crude approximation of the geometry of poles.

In part (a) of the figure, which is an aligned case, we see that both the poles perfectly overlap for the centred and off-centred case while they are at appreciable angles in the other two geometries as shown in (b) and (c). As the polar caps are linked to the location of poles, we get an estimate of the relative shift in the position of the polar cap for the centred and the off-centred geometry.

In (b) it is interesting to notice that $\theta$ for the poles i.e. the angle with respect to the $z$-axis, is smaller for both the poles in the off-centred case (blue) than in the centred case (red) which implies a shift in $\theta$ and might also be responsible for a small polar cap. In (c) when we look at $\phi$ i.e. the angle with respect to the $x$-axis, we witness that it is smaller in off-centred case than in the centred case for one pole but the opposite in the other, implying a sort of phase shift (discussed in Table~\ref{tab:pole_location}) in the location of poles which must be evident in the polar cap shapes too, which we discuss in the following paragraph.

Polar cap is the region mapped out by the foot points of magnetic field lines grazing the light cylinder. Considering studies, for instance, by \cite{rankin_toward_1993}, we believe that the pulsar radio emission is centred on the magnetic axis i.e. centred at the polar caps and hence, studying polar caps is crucial to understand the coherent radio emission. Polar caps strongly impact on the high-energy counterpart too, because slot gaps and the underlying two-pole caustic model of \cite{dyks_two-pole_2003} as well as outer gaps \citep{cheng_energetic_1986-2} also rely on the last closed field lines thus on the polar cap shape. We present a comparison of the polar cap geometry for the centred and the off-centred approach. To study the polar cap geometry we need to find the polar angle $\theta$ for the last closed field line i.e. the field line grazing the light cylinder. This approach serves as a fine estimation for the calculations associated with the polar caps. The procedure used for the calculations is described here in brief.

For a given $\phi$, a bracket for $\theta$ i.e. a range within which the $\theta$ corresponding to the last closed field line lies, is searched for. To find the last closed field line, for every field line corresponding to a $\theta$ lying between $0\degr$ and $90\degr$, a check is made if the field line reverses its direction when it encounters the light cylinder or not. If it doesn't, we move on to the next $\theta$ and repeat the check. When the check returns a value, another one close to it is checked for and saved so as to have a proper bracket to find a more precise value. Now, when we have a range within which our desired value surely exists, using the standard bisection method \citep{press_numerical_2007}, we find the best iterated solution for $\theta$, precise up to an accuracy of up to 10 decimal digits. These roots in $\theta$ corresponding to each $\phi$ are the points defining the polar cap boundary.

\begin{figure*}
   \centering
   \resizebox{0.95\textwidth}{!}{\input{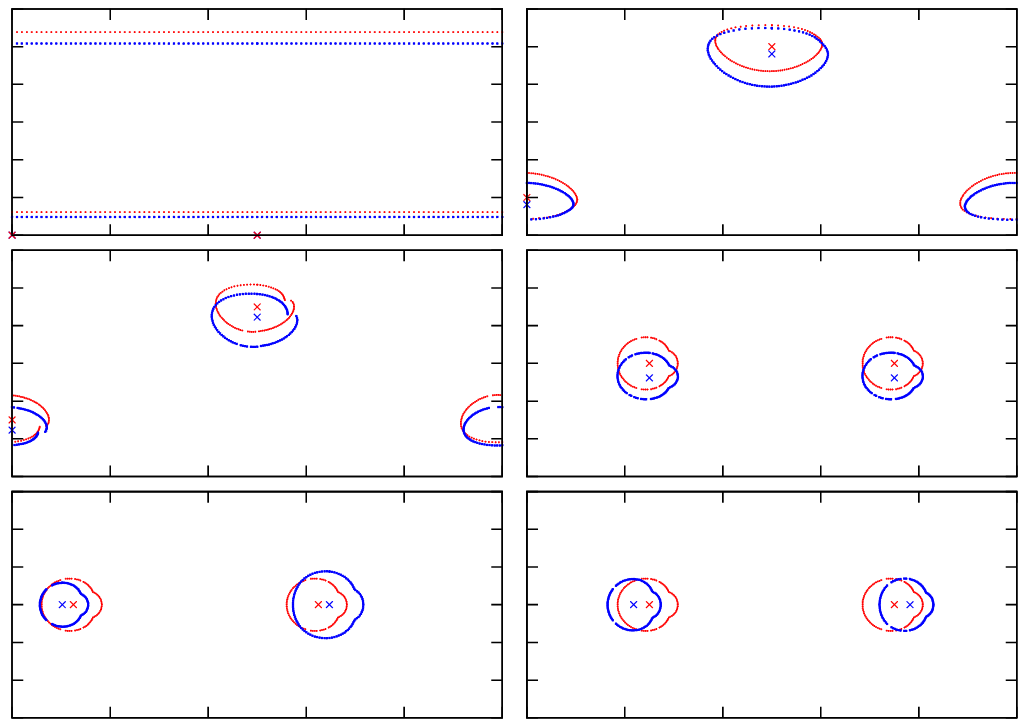}}
    \caption{Polar cap geometry comparison for centred case with $ \epsilon=0 $ (in red) and off-centred case with $ \epsilon=0.2 $ (in blue) for various cases. The $x$-axis and $y$-axis represent the azimuth $\phi$ and the polar angle $\theta$ respectively. Crosses are the location of the poles as calculated using Eqn.~(\ref{eq:mag_axis}).}
      \label{fig:polar_caps}
\end{figure*}

The shapes of the polar caps for the off-centred dipole are shown in Fig.~\ref{fig:polar_caps} for various inclination angles. $x$-axis and $y$-axis represent the azimuth angle $\phi$ and the polar angle $\theta$ respectively. The centred case (in red), shown for comparison, has been calculated by considering $ \epsilon = 0 $ and for the off-centred case (in blue), $ \epsilon $ is taken to be $ 0.2 $. The crosses represent the location of the poles with the colour scheme being the same as stated before.

We see that the polar caps are asymmetric and distorted as expected from a rotating dipole \citep{arendt_shape_1998}. A broad view distinction depicts that the shape is not affected due to the shift of the centre of the dipole from the geometrical centre but size and locations are, indeed, affected. The contrast in size is prominent for all the inclination geometries.

The size of the polar cap is one of the determining factors for the pulse profile width of pulsar. The larger or smaller polar caps as evident especially for $(\alpha, \beta, \delta) = (30\degr, 0\degr, 0\degr)$ and $(90\degr, 45\degr, 90\degr)$ could be used to justify observations with pulse widths showing discrepancy with respect to what is expected. For instance, as seen in the pulse width versus the rotation period plots in \cite{weltevrede_profile_2008}, there are many pulsars which lie away from the pulse width in the power law fit region. We know from such plots (also mentioned by \cite{kramer_characteristics_1998}, \cite{pilia_wide-band_2016}, \cite{malov_average_2010}) that there is a wide range of the pulse widths corresponding to almost all rotation periods possible, the extremes of which, larger or smaller widths, could be explained by this approach.

We refer to the geometries in which $\delta = 0\degr$ representing that the centred and the off-centred frames have their $z$-axis aligned with each other and the ones with $\delta = 90\degr$ representing that the two frames are perpendicular to each other as aligned and orthogonal cases respectively.

We show four aligned cases i.e.
$(\alpha, \beta, \delta) = (0\degr, 0\degr, 0\degr), (30\degr, 0\degr, 0\degr), (45\degr, 0\degr, 0\degr), (90\degr, 90\degr, 0\degr)$ in Fig.~\ref{fig:polar_caps} in which we clearly see the shift for the location of poles in $\zeta$. This is corresponded by the same variation which is prominent in Fig.~\ref{fig:pc_geom}(a) and (b) as discussed.

In contrast to the aligned case, the orthogonal geometries show a shift in $\phi$. The shift for each pole location is in opposite directions as seen in $(\alpha, \beta, \delta) = (90\degr, 45\degr, 90\degr)$ and $(90\degr, 90\degr, 90\degr)$. This variation in $\phi$ is also noticeable in Fig.~\ref{fig:pc_geom}(c).

Correspondingly, we see the shift for the polar caps in $\zeta$ for the aligned cases and in $\phi$ for the orthogonal cases. It has been noted by \cite{abdo_second_2013} in the second Fermi-LAT catalogue of gamma ray pulsars that pulsars indicate a phase lag between their radio and the high energy peak. The shift in $\phi$ (representing phase delay) for the orthogonal cases in our approach could be compared with observational phase delays in pulse profiles which could add to our understanding and hopefully, might serve as one of the factors, among many others like emission height, varying the retardation time by adjusting the emission altitude, aberration, magnetic field distortion, which are not yet fully known, to explain such observations.

\subsection{Error analysis}\label{Error analysis}
Expanding only up to the first order correction~$\epsilon$ in the electromagnetic field can be justified a posteriori by checking the error made by neglecting higher order terms. We expect errors of the order $\epsilon^2$ because higher order corrections require higher order multipoles. It was, indeed, shown by \cite{petri_radiation_2016} that an error of order $\epsilon^\ell$ is associated with neglecting multipoles starting at order $\ell+1$. In the present work, we consider up to $\ell=2$, the first correction. This discrepancy scaling like $\epsilon^2$ is tested in the following paragraph analysing the error made by computing the polar cap shape. To make such an analysis we compare the analytical solutions with the solutions we obtain from our numerical method. For simplicity, we consider only an aligned case with $(\alpha, \beta, \delta) = (0\degr, 0\degr, 0\degr)$.

To generate the analytical results for the aligned off-centred case, we shift the entire coordinate system under consideration for several values of the shift parameter $\epsilon$. In this shifted system the surface of the pulsar (with a shift $D$ along the $z$-axis) is represented by the equation
\begin{equation}
 x^{\prime2} + y^{\prime2} + (z^{\prime}+D)^2 = R^2
\end{equation}
where using the Cartesian to Spherical Coordinate System transformation we get,
\begin{equation}\label{eq:r_dash_quad}
 r^{\prime2} + 2D\cos{\theta^{\prime}}r^{\prime} + (D^2 - R^2) = 0.
\end{equation}
To find the analytical value of the root, we find the polar angle at which the field lines from this shifted pulsar intersects the light cylinder. For this, we need the equation of field lines grazing the light cylinder, $r^{\prime} = \rlight\sin^2{\theta^{\prime}} $ where $\rlight$ is the radius of the light cylinder. Substituting this condition in Eqn.~(\ref{eq:r_dash_quad}), we have:
\begin{equation}
\rlight^2 \, \sin^4 \theta' + 2\,D\,\rlight\,\sin^2\theta' \, \cos\theta' + D^2 - R^2 = 0.
\end{equation}
To calculate $\theta'$ from this equation, we consider an unknown $X=\cos\theta'$ and replace $\sin^2\theta'$ by $1-\cos^2\theta'$, and we get a quartic equation in $X$:
\begin{equation}\label{eq:X_quartic}
X^4 - 2\,\epsilon\,a \, X^3 - 2 \, X^2 + 2\,\epsilon\,a \, X + (\epsilon^2-1) \, a^2 + 1 = 0
\end{equation}
where we introduced $a=R/\rlight$. Solving this equation, we have $\theta'$ i.e. our required root but in off-centred frame. 

To be compared with numerical results, we need value of $\theta$ i.e. in rotation axis frame, therefore, a transformation is required. As it is an aligned case with shift along $z$-axis, the Cartesian transformation can be carried out simply using $ z = z' +D$. Using this and making transformation between Cartesian and spherical coordinate system using Appendix \ref{app:cc_sc}, we get analytical value of roots i.e. $\theta$.

\begin{table}
\centering
\begin{center}
\begin{tabular}{l l l l l l}
\toprule
\textbf{Shift} & \textbf{Analytical} & \textbf{Numerical} & \textbf{Error} & \textbf{Relative} & \boldmath{$R.E./\epsilon^2$}\\
\textbf{$\epsilon$} & \textbf{{value of $\theta$}} & \textbf{value of $\theta$} & \textbf{(in deg)} & \textbf{Error}\\
 & \textbf{(in deg)} & \textbf{(in deg)} &  & \textbf{$R.E.$} \\
\midrule
 & & \textbf{Pole 1}\\
0 & 18.4349 & 18.4349 & 0.0000 & 0.0000 & --\\
0.01 & 18.4349 & 18.1686 & 0.0048 & 0.0002 & 2.674\\
0.02 & 17.8937 & 17.9128 & 0.0191 & 0.0011 & 2.674\\
0.05 & 17.0906 & 17.2026 & 0.1121 & 0.0066 & 2.623\\
0.1 & 15.7762 & 16.1794 & 0.4032 & 0.0256 & 2.555\\
0.2 & 13.2434 & 14.5719 & 1.3285 & 0.1003 & 2.508\\
\midrule
 & & \textbf{Pole 2}\\
0 & 161.5650 & 161.5651 & 0.0000 & 0.0000 & --\\
0.01 & 161.2927 & 161.2876 & 0.0051 & 0.0000 & 0.316\\
0.02 & 161.0192 & 160.9982 & 0.0210 & 0.0001 & 0.327\\
0.05 & 160.1920 & 160.0499 & 0.1421 & 0.0009 & 0.355\\
0.1 & 158.7916 & 158.1424 & 0.6491 & 0.0041 & 0.409\\
0.2 & 155.9127 & 152.4586 & 3.4540 & 0.0222 & 0.554\\
\bottomrule
\end{tabular}
\end{center}
\caption{Error analysis for both poles with several shifts for aligned case $(\alpha, \beta, \delta) = (0\degr, 0\degr, 0\degr)$ with $R=0.1\rlight$.}
\label{tab:error_analysis}
\end{table}

We compare the error between the roots obtained from the analytical method explained above and our numerical approach in Table~\ref{tab:error_analysis}. The radius of pulsar $R$ is taken to be $0.1 R_{L}$.

The upper and lower section of the table shows the results for each of the poles. The left most column shows the shift parameter $\epsilon$ taken for a few values in between the centred case and going up to a shift of $20\%$ of $R$. It is clearly visible from the following two columns that for very small shifts the results are accurate up to second decimal place. However, we start to see the appreciable difference as we move towards larger shifts from the centre. These differences are highlighted in the next columns where we compute absolute error and the relative error.

It is significant to note that the order of the error is in conformity with what we expect by neglect of the higher order terms in the off-centred electromagnetic field equations. The equations have been considered to be precise up to an order of $\epsilon^2$ and we see the same order of accuracy while comparing the results from the analytical and the numerical approach. This, indeed, establishes that considering only a quadrupolar correction in the electromagnetic field equations is justified as long as $\epsilon$ remains small. If we go towards higher shifts, error becomes larger scaling as $\epsilon^2$. If we accept truncation errors up to several percent, let us fix the limit to~5\%, then adding only the quadrupolar corrections to the dipole is justified because the first neglected term, the hexapolar component $\ell=3$, is of the order $\epsilon^2=0.04$ $(4\%)$. Going to $\epsilon\gtrsim0.25$ would lead to too high truncation errors impacting on the precision of the pulse profile unacceptable with the high quality of observations from radio to gamma-rays. Adding higher order multipoles is the only way to study satisfactorily larger off-centred dipoles.

We can write the magnetic field equation in simple terms up to second order in $\epsilon$ as $B = B_{dip} + \epsilon B_{quad} + \epsilon^{2} B_{hex} $ or, for more convenience, as $B - B_{dip} - \epsilon B_{quad} = \epsilon^{2} B_{hex} $. $B$ is the total magnetic field while $B_{dip}$, $B_{quad}$ and $B_{hex} $ are the dipolar, quadrupolar and hexapolar representatives of the same. As the left hand side of this equation is giving us an estimate of error between the total magnetic field and magnetic field put as a sum of its components, we expect that the ratio of relative error to $\epsilon^{2}$ will give us a good estimate of the component $B_{hex} $. As we would expect that this value should be a constant, we see in the last column of Table~\ref{tab:error_analysis} that they are quite close for all the cases; at least up to first decimal place. This provides an even stronger claim for the point that the equations are accurate up to second order in $\epsilon$.

\subsection{Radio and high energy phase plots}\label{Radio and high energy phase plots}
Observing light curves in the radio and high energy bands offers a mean to diagnose the magnetic topology inside the magnetosphere; close to the surface for radio pulse profiles and close to the light-cylinder for high-energy light curves. Accurate fits of broad band observations gives us some hints about the possible deviation from a perfect centred dipole. In this section, we present detailed phase plots for radio and high energy emission. 

The two-pole caustic model for calculating high energy emission light curves as explained by \cite{dyks_two-pole_2003} is used. However, the aberration formula used by them was justified as only a reasonable approximation by \cite{bai_uncertainties_2010}. We will be presenting both of them to show the differences for our off-centred approach. The former used the following aberration formula to transform photon propagation direction from the corotating frame $\hat{\boldsymbol{\eta}'}$ to the inertial frame $\hat{\boldsymbol{\eta}}$ of a distant observer.
\begin{equation}\label{eq:aberration}
\hat{\boldsymbol{\eta}} = \frac{\hat{\boldsymbol{\eta}'}+(\gamma+(\gamma-1)(\boldsymbol{\beta_{c}} \cdot \hat{\boldsymbol{\eta}'}) / \boldsymbol{\beta_{c}^{2}}) \boldsymbol{\beta_{c}}}{\gamma(1+\boldsymbol{\beta_{c}} \cdot \hat{\boldsymbol{\eta}'})}
\end{equation}
where $\gamma = 1/\sqrt{(1-\boldsymbol{\beta_{c}^{2}})}$ and $\boldsymbol{\beta_{c}} = (\boldsymbol{\omega} \times \mathbf{r})/c$ with $\boldsymbol{\omega}$ being the angular velocity of the pulsar and $\mathbf{r}$ being the radial position of the emission point. The revised version given by the latter taking photon propagation direction in the corotating frame as $\boldsymbol{\beta_{c}}$ and in the lab frame as $\boldsymbol{\beta_{0}}$, was given by;
\begin{equation}\label{eq:aberration_revised}
\boldsymbol{\beta_{0}} = f \mathbf{B} + \boldsymbol{\beta_{c}}
\end{equation}
where $f$ is a coefficient which is determined by implementing the $\abs{\boldsymbol{\beta_{0}}}\rightarrow 1$ condition for the emitted particle. Also, in this limit, $\boldsymbol{\beta_{0}} \rightarrow \hat{\boldsymbol{\eta}}$. The equation gives two solutions; one corresponds to the particle emitted in the direction of the magnetic field line while the other corresponds to the opposite direction. The former solution i.e. the forward moving particle is to be considered.

Also, to consider the photon travel delays, the phase $\phi$ is taken to be
\begin{equation}\label{eq:photon_delay}
\phi = -\phi_{em} - \mathbf{r} \cdot \hat{\boldsymbol{\eta}}/R_{L}
\end{equation}
where $\phi_{em}$ is the azimuth for the direction $\hat{\boldsymbol{\eta}}$ and $R_{L}$ is the radius of the light cylinder.

\begin{figure}
   \centering
   \resizebox{0.95\textwidth}{!}{\input{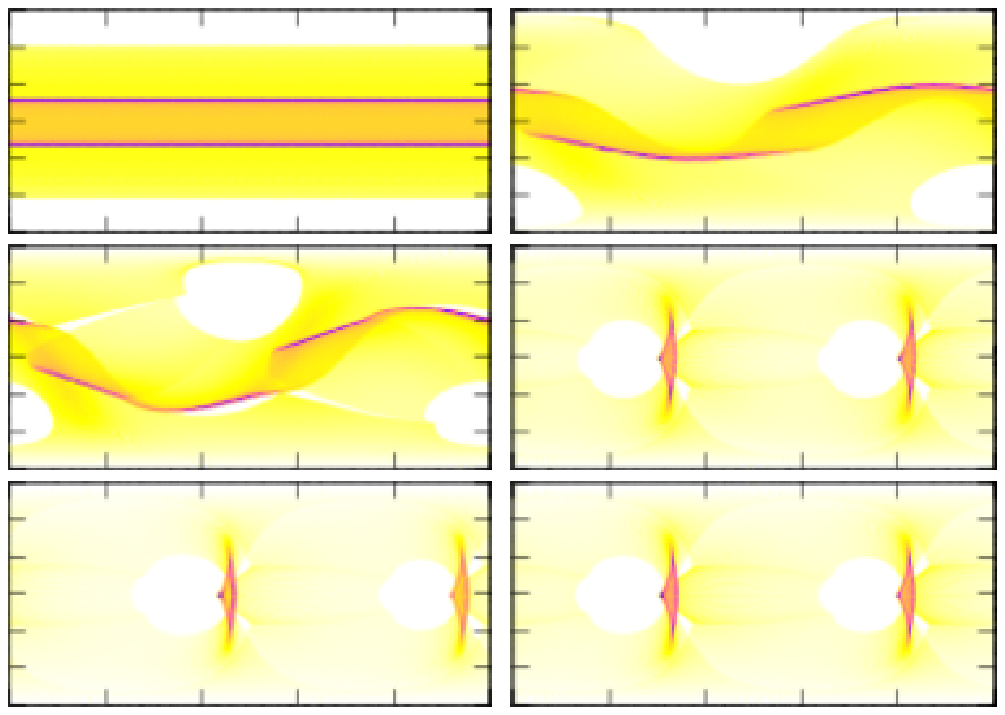}}
    \caption{High energy emission phase diagrams for off-centred case with $ \epsilon = 0.2 $ for various cases for the aberration formula as given by \protect\cite{dyks_two-pole_2003}. The white region corresponds to region with no emission and moving towards blue-black via yellow-orange signifies an increase in the photon count.}
      \label{fig:he_phase_diagrams_dyks_aberration}
\end{figure}

\begin{figure}
   \centering
   \resizebox{0.95\textwidth}{!}{\input{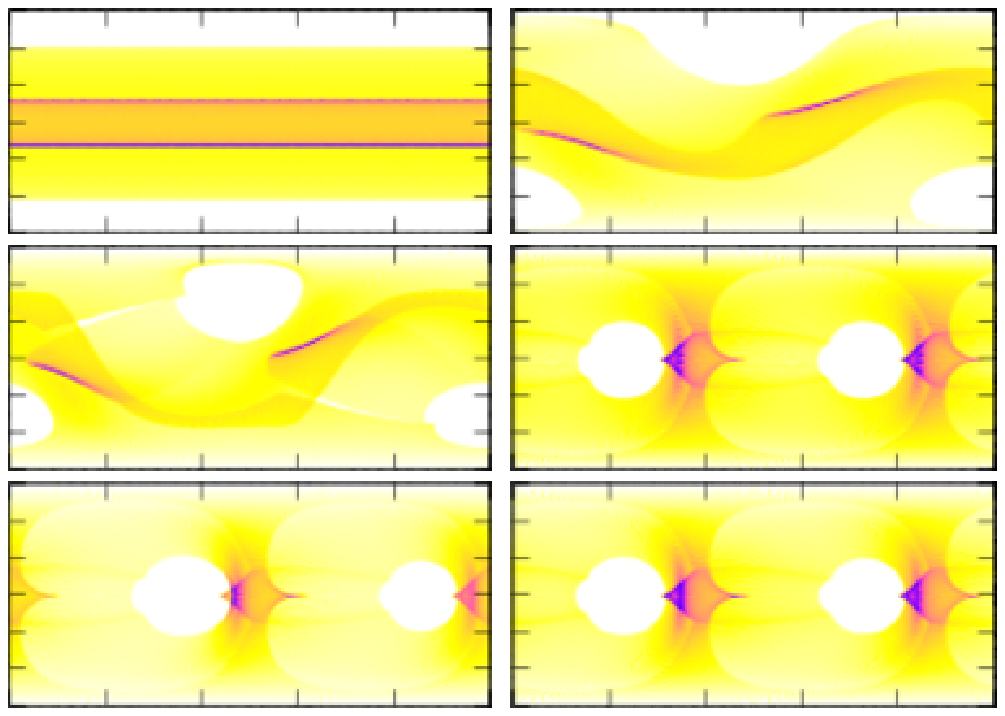}}
    \caption{High energy emission phase diagrams for off-centred case with $ \epsilon = 0.2 $ for various cases for the aberration formula corrected by \protect\cite{bai_uncertainties_2010}. The white region corresponds to region with no emission and moving towards blue-black via yellow-orange signifies an increase in the photon count.}
      \label{fig:he_phase_diagrams}
\end{figure}

We show the high energy emission for $ \epsilon = 0.2 $ for several cases using above formulae. Fig.~\ref{fig:he_phase_diagrams_dyks_aberration} uses  Eqn.~(\ref{eq:aberration}) and Fig.~\ref{fig:he_phase_diagrams} uses the revised Eqn.~(\ref{eq:aberration_revised}). In both figures, the $x$-axis spans the phase $ \phi $ while the $y$-axis represents the angle of the line of sight $\zeta$. The white region corresponds to 'no emission' region and moving towards blue-black via yellow-orange signifies an increase in the photon count.

Due to caustic effects, in the two-pole caustic model, significant emission emanates from regions close to the light-cylinder. We see these high energy emission regions in blue-black in all the cases. The white regions specify no emission observed when the line of sight crosses the polar caps which are free of high energy photons. Also, the asymmetry in the size of the polar cap for the two poles as seen in Fig.~\ref{fig:polar_caps} for $(\alpha, \beta, \delta) = (90\degr, 45\degr, 90\degr)$ is reflected in the emission too. It is important to note that there is a small 'bump' disturbing the symmetry of the polar cap shape in Fig.~\ref{fig:polar_caps} which is also visible in Fig.~\ref{fig:he_phase_diagrams_dyks_aberration} and Fig.~\ref{fig:he_phase_diagrams} but the direction stands reversed. This is because these two plots describe totally different scenarios. The former is merely a geometrical description of the polar cap while the latter two are representing the photon emission at respective points, and hence, the reflected asymmetry is not something surprising.

It was pointed out by \cite{bai_uncertainties_2010} for the centred case that the corrected aberration leads to wider caustics. We notice the same trend to be followed in the off-centred case as the corrected aberration in Fig.~\ref{fig:he_phase_diagrams} displays a different caustic structure than the older version in Fig.~\ref{fig:he_phase_diagrams_dyks_aberration}. We will use the corrected formulae for the radio emission later without showing it separately for both cases because aberration effects are not significant in that region.

\begin{figure}
   \centering
   \resizebox{0.95\textwidth}{!}{\input{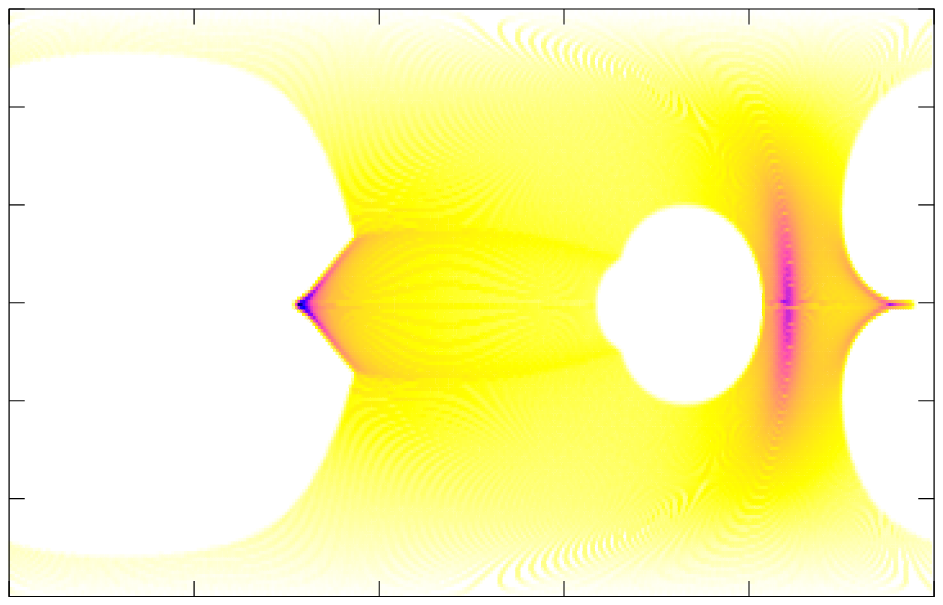}}
    \caption{High energy emission phase diagram for an off-centred ($ \epsilon = 0.2 $) case of $(\alpha, \beta, \delta) = (90\degr, 90\degr, 90\degr)$ for one pole. The white region corresponds to region with no emission and moving towards blue-black via yellow-orange signifies an increase in the photon count.}
      \label{fig:phase_90_one_pole}
\end{figure}

Our calculations for the high energy emission phase diagrams for the centred case ($ \epsilon = 0$) given by Eqn.~(\ref{eq:aberration}) and those given by Eqn.~(\ref{eq:aberration_revised}) match well with \cite{petri_multipolar_2015} and \cite{bai_uncertainties_2010} respectively. The small differences are only because the boundary conditions used by us are different from earlier works; we terminate our calculations of the photon count at a distance going as far as $0.95 \rlight$ (other works used $0.75-0.80 \rlight$). Also, in orthogonal cases with $\alpha = 90\degr$, we notice sharp peak points close to both the poles. This is because of our consideration of farther boundary because of which the emission from pole $1$, calculated separately, extends up to the phase corresponding to pole $2$ and hence, adds to the emission of pole $2$. This behaviour is highlighted in Fig.~\ref{fig:phase_90_one_pole} where result for one pole (on the right; say pole $1$) is shown. Calculation of photon count only from pole $1$ shows it extending up to the phase of pole $2$. Similar behaviour is observed from individual calculations for pole $2$. This justifies the sharp peak points noticed in the orthogonal cases.

It is difficult to precisely conclude anything related to the off-centred approach from these phase diagrams because the differences are small. Thus, in the next section we pick out some special geometric configurations to point out the major discrepancies between centred and off-centred dipole but before showing these pulse profiles and light-curves, let's have a look at the phase diagrams for the radio emission.

Current wisdom assumes that radio emission is produced mostly in the vicinity of the polar caps up to altitude of several hundredths of kilometres. These conclusions are drawn from polarization angle studies conducted by several authors. See for instance \citet{mitra_comparing_2004}, \citet{gangadhara_understanding_2001} and \citet{johnston_evidence_2007} where the shift in the inflexion point of polarization is used to localize the emission site. This statement is generally not true for millisecond pulsars or for the Crab which possess extended collocated radio and high-energy emission and some phase-aligned from radio through optical, X-ray up to gamma-ray pulses \citep{moffett_multifrequency_1996}. We focus on this radio emission region by specifying a minimum altitude~$H_l$ and a maximum altitude $H_u$ where emission is produced. These distances are measured in spherical coordinates such that the volume of emission is comprised between the radii $r=H_l$ and $r=H_u$.

\begin{figure}
   \centering
   \resizebox{0.95\textwidth}{!}{\input{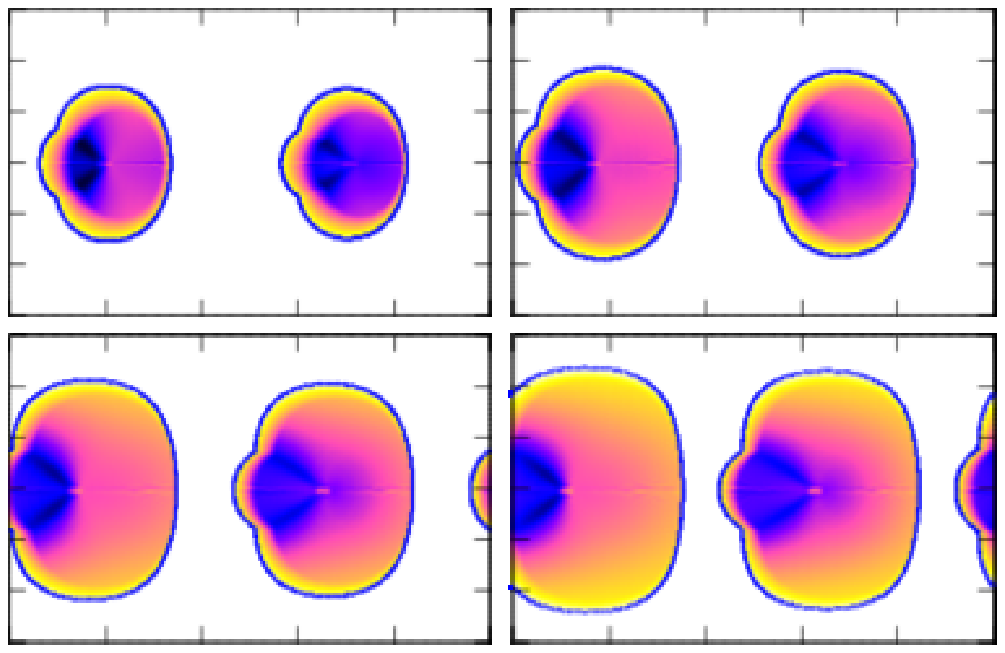}}
    \caption{Radio emission phase diagrams for several sections of heights between $[H_l, H_u] = [R, 5R]$ for $ (\alpha, \beta, \delta) = (90\degr, 90\degr, 90\degr)$ for an off-centred case with $ \epsilon = 0.2 $. The blue boundary depicts the outer rim of the emission region. The white region corresponds to region with no emission and moving towards blue-black via yellow-pink signifies an increase in the photon count.}
      \label{fig:re_phase_rim_heights_D20}
\end{figure}

Before analysing the radio emission phase diagrams for different cases, we show how the emission varies with respect to the height from the surface for a single case of $(\alpha, \beta, \delta) = (90\degr, 90\degr, 90\degr)$ for $ \epsilon = 0.2 $ in Fig.~\ref{fig:re_phase_rim_heights_D20}. We consider four distinct emission volumes, each of height one stellar radius~$R$, going from the surface up to altitude which is four times $R$ and look at the emission from those regions. More explicitly, the intervals chosen are $[R,2\,R]$, $[2\,R,3\,R]$, $[3\,R,4\,R]$ and $[4\,R,5\,R]$. To highlight the boundary of the emission region, we show the rim of each region as a surrounding perimeter in blue. The white region corresponds to region with no emission and moving towards blue-black via yellow-pink signifies an increase in the photon count.

\begin{figure}
   \centering
   \resizebox{0.95\textwidth}{!}{\input{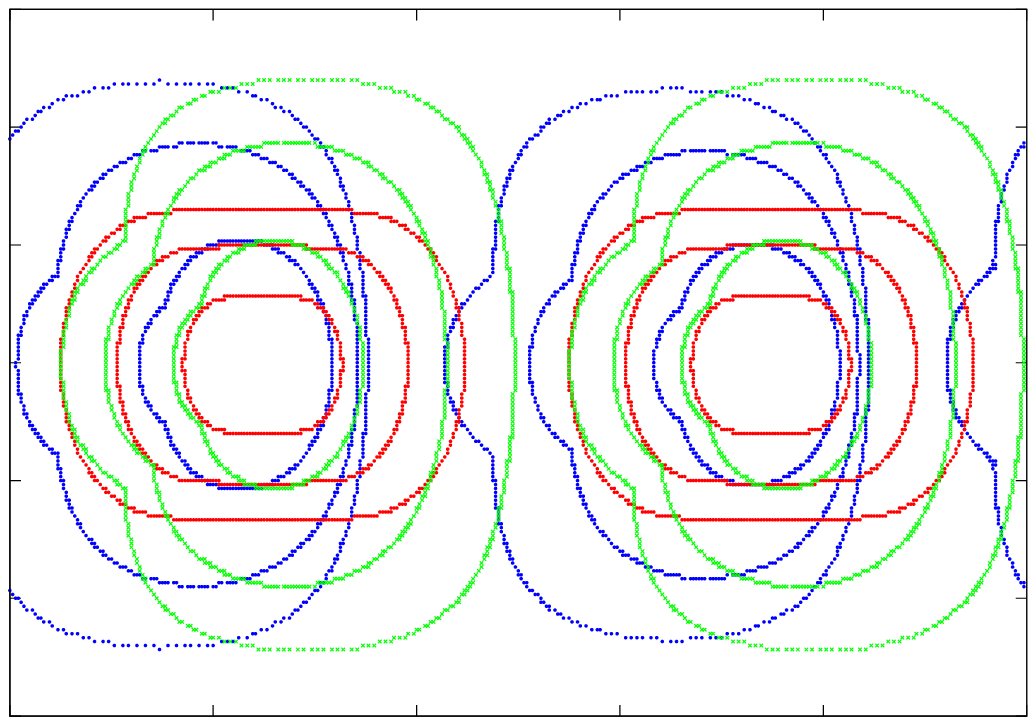}}
    \caption{Outer rims of the radio emission region for heights $R, 3R, 5R$ (radial distance from centre) for $ (\alpha, \beta, \delta) = (90\degr, 90\degr, 90\degr)$. The blue rims are for an off-centred case ($ \epsilon = 0.2 $) with retardation and aberration effects included, green rims for Deutsch solutions without accounting for retardation and aberration while the red rims correspond to a static centred dipole. The smallest rim correspond to $H_l = R$ and the largest one to $H_u = 5R$. The intermediate rims have been removed for a clear visualisation.}
      \label{fig:re_rim_off_deutsch_static}
\end{figure}

Clearly, the radio emission is in the vicinity of the polar caps and is strongest when considered close to the centre. The emission decreases in intensity with an increase in altitude. This is because higher the altitude, more sparse the magnetic field lines and lesser is the photon density as they are spread out over a larger solid angle now. For a better understanding, we compare this off-centred case with a simple case of static dipole and also with Deutsch field solutions in Fig.~\ref{fig:re_rim_off_deutsch_static} where we plot the rims of the radio emission region at several heights for all three cases. The red rims are for the static centred dipole case, green rims for Deutsch solutions without accounting for retardation and aberration while blue rims represent the boundary of emission region for an off-centred case with retardation and aberration effects included. Starting from $H_l = R$ and going until $H_u = 5R$, we notice an increase in size of the rims. Also, for static dipole the rims are symmetric but as soon as we include rotation (Deutsch solutions), an asymmetry is introduced which is further intensified on inclusion of retardation effects in the off-centred case.

As is clear from the figure, the width of the rim, which is basically an indication of the width of the pulse, increases with an increase in height. This relation between the altitude of emission and pulse width can be easily calculated for a simple orthogonal static centred dipole using following method.

To simplify our discussion, let us assume that photons are emitted in the equatorial plane. As this plane is a symmetry plane, photons are restricted to move in this plane. We set up a polar coordinate system $(r,\phi)$. The last closed field line is then represented by $r=\rlight\,\sin^2\phi = R \, \sin^2\phi/\sin^2\phi_{\rm pc}$. Photons are emitted in a direction tangent locally to field lines. This tangent vector is defined by
\begin{equation}
\mathbf{t} = \frac{d\mathbf{r}}{ds}
\end{equation}
where $ds^2=\rlight^2\,\sin^2\,\phi\,(3\,\cos^2\phi+1)\,d\phi^2$ is the infinitesimal curvilinear abscissa along the last closed field line. Expressing the tangent vector in Cartesian coordinates, the angle of the photon direction with respect to the $x$-axis becomes
\begin{equation}
\tan \alpha = \frac{t_y}{t_x} = \frac{3\,\cos\phi\,\sin\phi}{3\,\cos^2\phi-1}.
\end{equation}
The altitude of emission~$H=\eta\,R$ is related to the angle~$\phi$ by $H=\rlight\,\sin^2\phi$ and thus the photon created at a height~$H$ is pointing into a direction
\begin{equation}
\tan \alpha_H = \frac{\sqrt{\frac{H}{\rlight} \, \left(1-\frac{H}{\rlight}\right)}}{ \frac{2}{3} - \frac{H}{\rlight} } = \frac{ \sqrt{\eta\,a \, ( 1 - \eta\,a)}}{ \frac{2}{3} - \eta\,a} 
\end{equation}
where $a = \frac{R}{R_{L}}$.

A good estimate for the pulse width at an altitude~$H$ is therefore $w=2\,\alpha_H$ or in fraction of the total period~$\alpha_H/\upi$. If the magnetic moment points towards the observer at time $t=0$ then the radio pulse will be seen starting from time $t=-\alpha_H/\Omega$ up to time $t=\alpha_H/\Omega$.

We make an estimate of the widths using this formula for several values of $\eta$ and compare it with the numerical values in Table~\ref{tab:pulse_width}. All the numerical calculations for the phase plots have been made considering a resolution of $1\degr$ which limits the calculation of pulse widths to the same degree of accuracy. The analytical values are in an excellent agreement with the numerical values for a static centred dipole. We also show the corresponding values for the Deutsch solutions and for an off-centred case with $\epsilon=0.2$. The width remains same for both the poles except in the off-centred case as shown in separate columns where retardation and aberration effects are under consideration.

\begin{table}
\centering
\begin{center}
\begin{tabular}{l l l l l l}
\toprule
\boldmath{$\eta$} & \textbf{Static dipole} & \textbf{Static dipole}& \textbf{Deutsch} & \textbf{Off-centred} & (\boldmath{$\epsilon=0.2$}) \\
 & \textbf{(Analytical)} & \textbf{(Numerical)} & \textbf{(Numerical)} & \textbf{(Numerical)} \\
 & & & & \textbf{Pole 1} & \textbf{Pole 2}\\
\midrule
1 & $55.8\degr$ & $57\degr$ & $67\degr$ & $66\degr$ & $68\degr$\\
2 & $81.2\degr$ & $81\degr$ & $97\degr$ & $95\degr$ & $99\degr$\\
3 & $102.7\degr$ & $103\degr$ & $121\degr$ & $117\degr$ & $121\degr$\\
4 & $122.9\degr$ & $123\degr$ & $142\degr$ & $134\degr$ & $139\degr$\\
5 & $143.1\degr$ & $143\degr$ & $161\degr$ & $150\degr$ & $155\degr$\\
\bottomrule
\end{tabular}
\end{center}
\caption{Pulse widths for various values of $\eta=H/R$ for an orthogonal static centred dipole (analytical and numerical both), for Deutsch fields (numerical) and for both poles an off-centred case with $\epsilon=0.2$ (numerical). The width is same for both the poles in the former two cases and hence, not mentioned separately. The retardation and aberration effects are considered only in the latter case.}
\label{tab:pulse_width}
\end{table}

There is another important point to be noticed in Fig.~\ref{fig:re_phase_rim_heights_D20}. We notice a shift in the entire emission region which moves towards lower $\phi$ as we move towards higher heights. For instance, the phase for the interval $[4\,R,5\,R]$ is shifted to left (lower $\phi$) as compared to that in $[R,2\,R]$. This shift is also prominent in Fig.~\ref{fig:re_rim_off_deutsch_static} but only for the retarded off-centred case. For the static dipole case without any retardation, the geometrical centre of the rim is observed to remain fixed for all heights. This indicates that this shift in phase with heights is affiliated to retardation effects because of the term $- \mathbf{r} \cdot \hat{\boldsymbol{\eta}}/R_{L}$ discussed in Eqn.~(\ref{eq:photon_delay}).

\begin{figure}
  \centering
   \begin{tabular}{c}
   \includegraphics[width=0.75\textwidth]{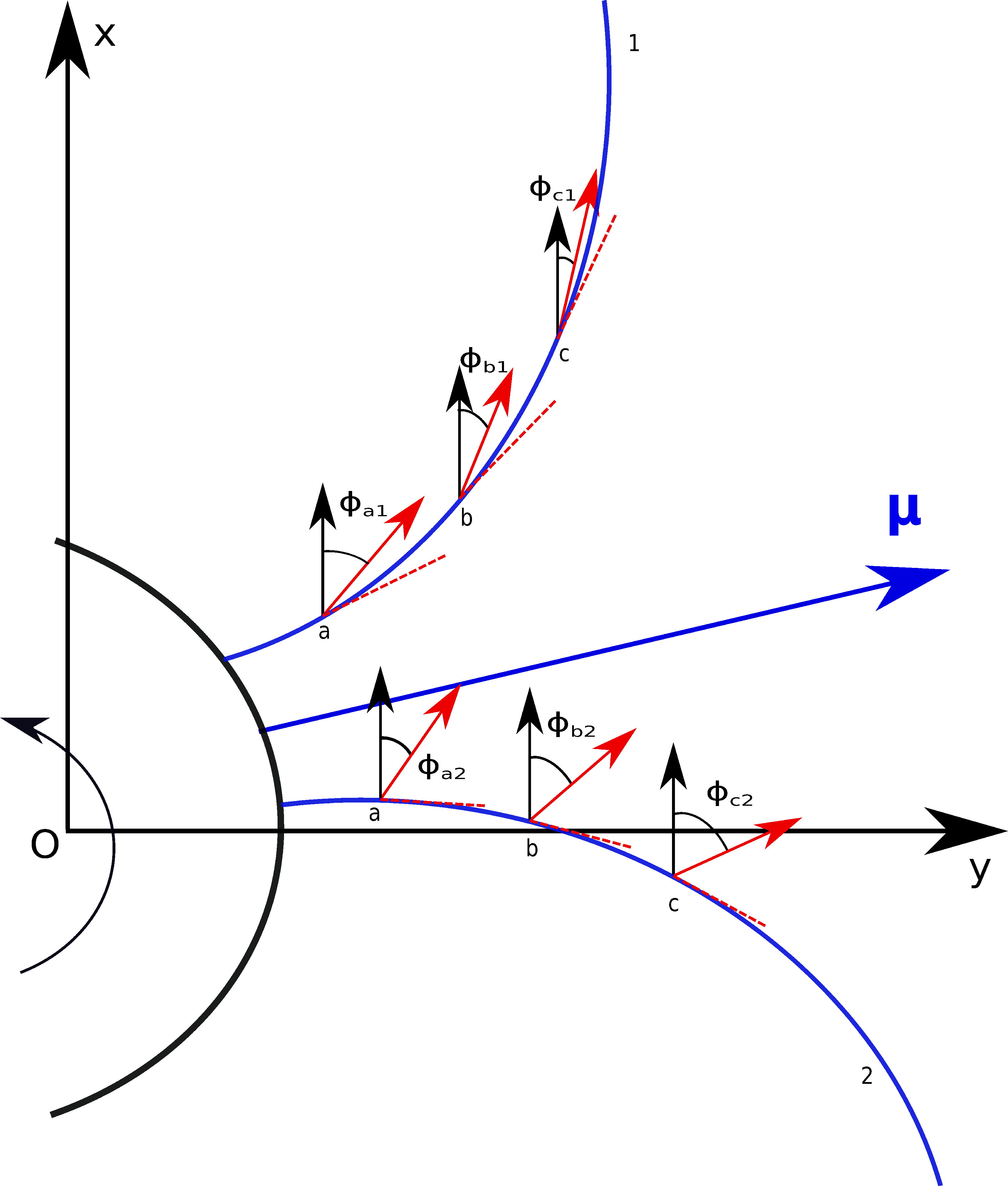}
  \end{tabular}
  \caption{2D representation of the equatorial plane for the off-centred geometry with $ (\alpha, \beta, \delta) = (90\degr,90\degr,90\degr)$ to show the phase of the photon propagation direction. $z$-axis is pointing in the plane of paper. $O$ is the geometrical centre of the pulsar, a part of whose surface is shown as black solid line. The curved black arrow represents the anti clockwise direction of the rotation of the pulsar. Blue vector is the magnetic moment and blue solid curved lines are magnetic field lines; the one labelled as $1$ is the field line corresponding to leading peak while $2$ is the same for the trailing peak. $a$, $b$ and $c$ are the three points on the field lines in increasing order of heights (radial distance from the geometric centre). The red dashed line represents the direction of the photon propagation at any instant without any retardation under effect while red solid arrow represents the direction of the path followed by photon as seen by an observer when retardation effects are included. $\phi$ is an indication of the phase for the photon propagation with subscripts describing the point and the field line. Figure not to scale and provide only a rough depiction.}
  \label{fig:photon_path}
\end{figure}

This behaviour is explained geometrically in Fig.~\ref{fig:photon_path} \citep{dyks_rotation_2002}. The figure is not to scale and is only used for a rough approximation. In the figure, a 2D representation of the equatorial plane is shown with $z$-axis pointing in the plane of paper. $O$ is the geometrical centre of the pulsar, a part of whose surface is shown in black solid line. The curved black arrow represents the anti clockwise direction of the rotation of the pulsar. Blue vector is the direction of the magnetic moment (not necessarily perpendicular to the surface as shown) and blue solid curved lines are two magnetic field lines under consideration. The magnetic field line which is labelled as $1$ is responsible for the leading peak of emission while $2$ is the corresponding trailing emission peak.

Now, we study the path of the photon propagation along these two field lines. We consider photons at three different points $a$, $b$ and $c$ on the field lines in increasing order of heights (radial distance from the geometric centre). Without aberration, the red dashed line is the direction of the photon propagation which is nothing but the tangent at the point of consideration. Now, as the pulsar rotates, the aberration effects influence the direction of photon travel as observed by an observer. Considering the observer is looking in the $x$ direction (shown in small black solid arrows), the photon propagation in an observer's frame is shown as red solid arrows. The observer tend to observe the photon a bit earlier because of the retardation coming into play as the pulsar rotates.

The angle $\phi$ represents the phase of the photon with first subscript indicating the point and second subscript mentioning the field line considered. For instance, for the leading peak, the photon at point $a$ has phase ${\phi}_{a1}$. As evident from the figure geometry, ${\phi}_{a1} > {\phi}_{b1} > {\phi}_{c1}$ which implies that as we move higher, the phase $\phi$ decreases which justifies the shift observed towards lower $\phi$ with increasing height in Fig.~\ref{fig:re_phase_rim_heights_D20} and also in Fig.~\ref{fig:re_rim_off_deutsch_static} for the leading peak. And, for the trailing peak, we notice the opposite i.e. ${\phi}_{a2} < {\phi}_{b2} < {\phi}_{c2}$ , which is apparent in Fig.~\ref{fig:re_rim_off_deutsch_static} where the trailing edge of emission has higher $\phi$ for higher heights.

Next we consider the full emission volume starting from $ H_l = R $ upto $ H_u = 5R $ for calculating the radio emission phase diagrams.

\begin{figure}
   \centering
   \resizebox{0.95\textwidth}{!}{\input{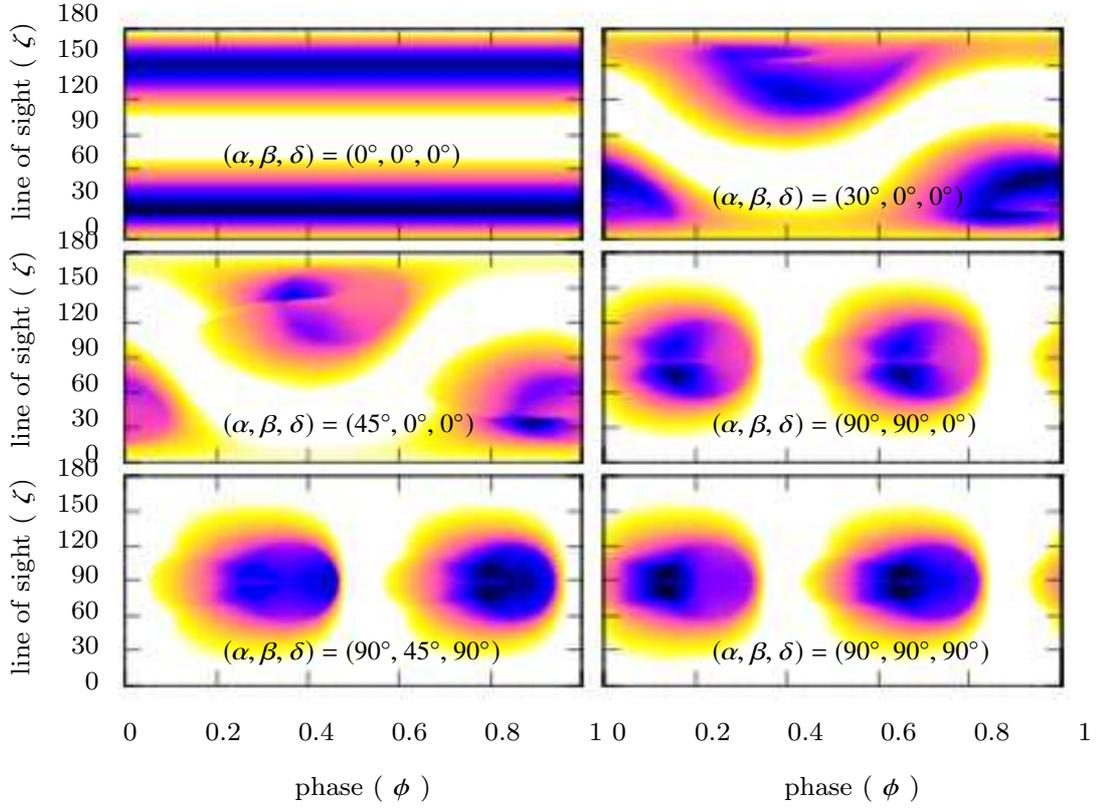}}
    \caption{Radio emission phase diagrams for off-centred case with $ \epsilon = 0.2 $ for various cases. The white region corresponds to region with no emission and moving towards blue-black via yellow-orange signifies an increase in the photon count.}
      \label{fig:re_phase_diagrams}
\end{figure}

Fig.~\ref{fig:re_phase_diagrams} presents the phase diagram for the radio emission for the same parameters used as for the high energy emission in Fig.~\ref{fig:he_phase_diagrams}. There are some discontinuities which are quite apparent in some cases like a strange line in the equatorial region for the orthogonal case. It persists for several different samplings considered to calculate the emission. However, we intend to apply more uniform sampling in future to improve the results even further.

We see the emission regions clearly complement the high energy emission regions that is emission is almost only within the polar caps contrary to high energy. We have considered emission starting from the surface of the pulsar going to a height which is five times its radius. The emission slowly decreases in intensity as we move away from the centre of the polar cap because photons are spread in wider directions when moving away from the polar cap centre. It is interesting to notice that the emission is not identical for the two poles which is affiliated to the retardation effects. However, aberration can usually be neglected for photons produced in the vicinity of the neutron star surface.

\subsection{Radio and high energy emission light curves}\label{Radio and high energy emission light curves}
Two dimensional phase plots give full information about the emission patterns but they render difficult direct comparison between centred and off-centred cases. In order to better assert the phase shift between both geometries, we extract specific light-curves and radio pulse profiles overlapping both topologies on a same graph.

To be able to do so, we need to choose a particular line of sight to see how the photon intensity varies over the whole range of $\phi$ with it. The intensities have been normalized to unity for an easy comparison.

\begin{figure}
   \centering
      \begin{tabular}{c}
      \subfloat[ $~\zeta = 20\degr~,30 \degr~,40 \degr~,50 \degr$]{\resizebox{!}{0.6\textwidth}{\input{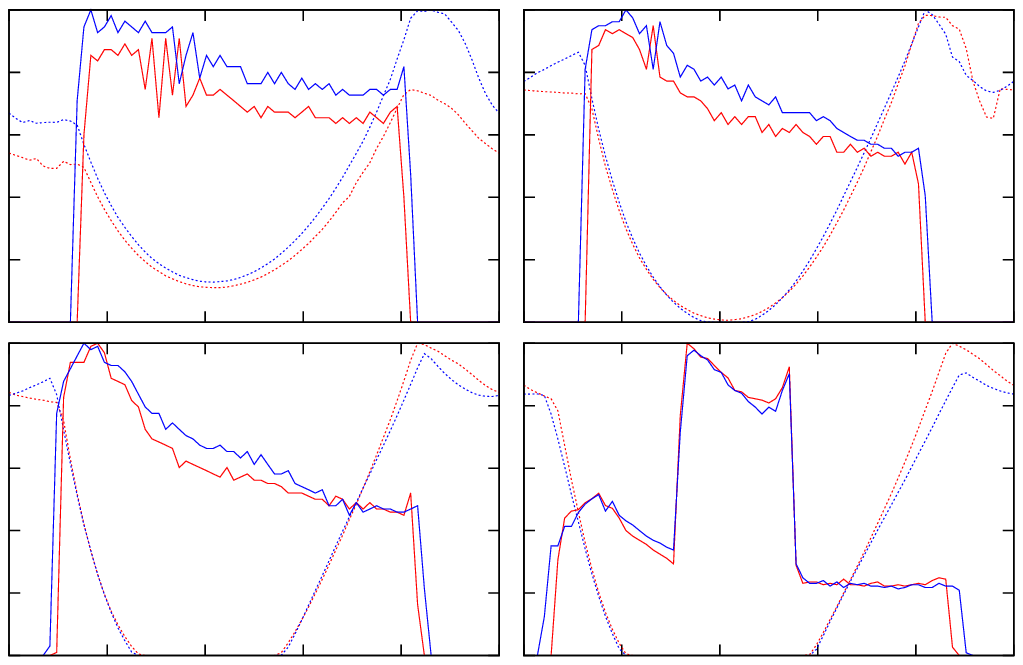}}}\\
      \subfloat[ $~\zeta = 130\degr~,140 \degr~,150 \degr~,160 \degr$]{\resizebox{!}{0.6\textwidth}{\input{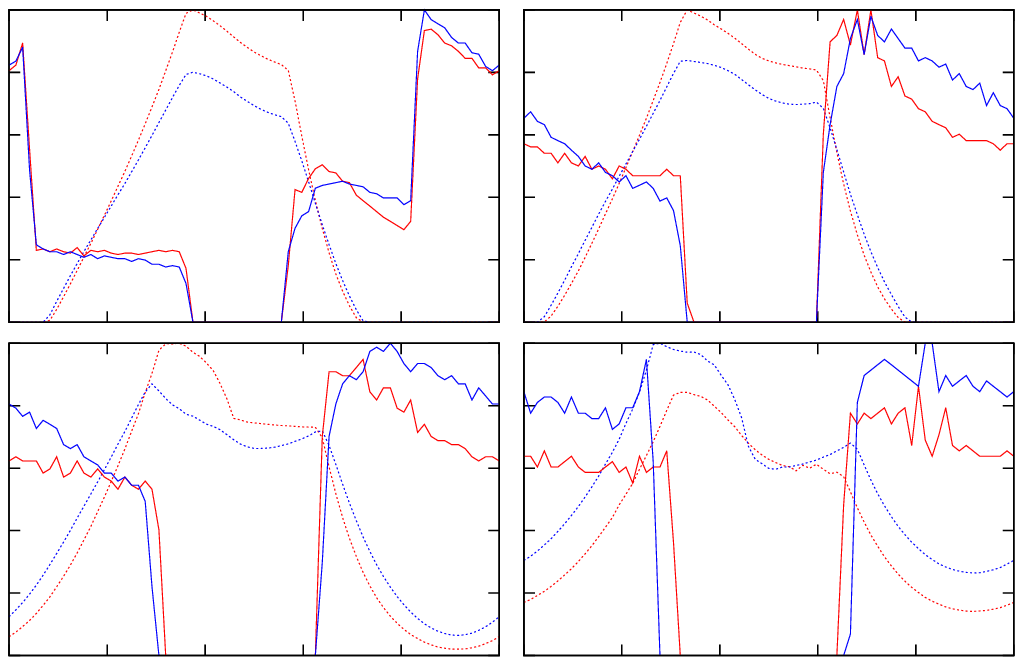}}}    
      \end{tabular}
    \caption{High energy and radio emission light curves for $ (\alpha, \beta, \delta) = (30\degr, 0\degr, 0\degr)$ for centred case ($ \epsilon = 0 $, in red) and off-centred case ($ \epsilon = 0.2 $, in blue) for various line of sights ($\zeta$). In (a) we have $\zeta$ ranging from $20\degr$ to $50\degr$ and (b) has $\zeta$ from $130\degr$ to $160\degr$ in steps of $10\degr$ each. The high energy emission curves are represented in solid lines and the radio emission curves in dashed lines.}
      \label{fig:off_lc_30_0_0}
\end{figure}

\begin{figure}
   \centering
    \resizebox{0.95\textwidth}{!}{\input{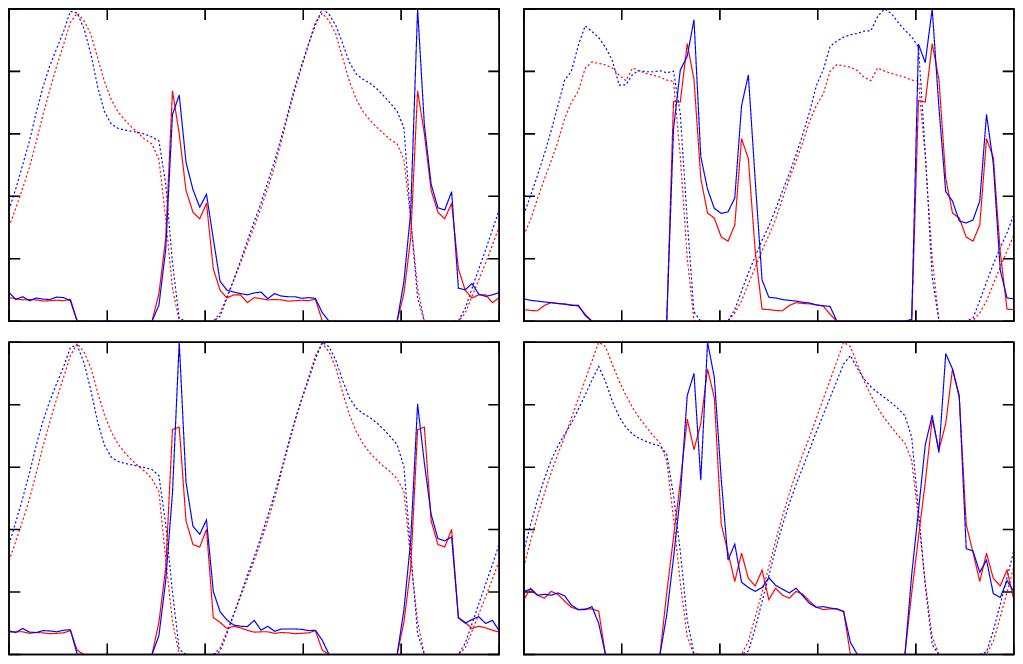}}
    \caption{High energy and radio emission light curves for $ (\alpha, \beta, \delta) = (90\degr, 90\degr, 90\degr)$ for centred case ($ \epsilon = 0 $, in red) and off-centred case ($ \epsilon = 0.2 $, in blue) for various line of sights ($~\zeta = 80\degr~,90 \degr~,100 \degr~,110 \degr$). The high energy emission curves are represented in solid lines and the radio emission curves in dashed lines.}
      \label{fig:off_lc_90_90_90}
\end{figure}

The light curves for high energy and radio regime for $(\alpha, \beta, \delta) = (30\degr, 0\degr, 0\degr)$ with $ \epsilon = 0 $ (in red) and $ \epsilon = 0.2 $ (in blue) for various line of sights ($\zeta$) are shown in Fig.~\ref{fig:off_lc_30_0_0}. The angles of line of sight $ \zeta $ are chosen in a range between $20\degr$ and $50\degr$ in (a) and between $130\degr$ and $160\degr$ in (b) in steps of $10\degr$ each. We also show another case with $(\alpha, \beta, \delta) = (90\degr, 90\degr, 90\degr)$ in Fig.~\ref{fig:off_lc_90_90_90} with $\zeta$ ranging between $80\degr$ to $110\degr$. The high energy emission curves are represented in solid lines and the radio emission curves in dashed lines. All the curves are normalized in terms of intensity. The normalization factor chosen for the radio and high energy emission is the maximum photon count corresponding to each of those cases separately.

A quick look at Fig.~\ref{fig:off_lc_30_0_0} reveals that the radio emission and the high energy emission complement each other. The former peaks in the vicinity of the polar regions which is in accord with the current models based on the assumption that radio emission is produced exclusively near the polar caps while the latter peaks where radio emission is at its lowest.

For $\zeta = 20\degr$, radio emission peak for the off-centred case is higher in intensity than that for the centred case. As we move towards higher $\zeta$ in Fig.~\ref{fig:off_lc_30_0_0}(a) we notice there is a switch between them. The fact that the radio emission peaks at a lower $\zeta$ for the off-centred case while at a higher value for the centred case actually denotes a shift in the polar cap as the emission region for radio regime lies in the proximity of the polar caps. This highlights the shift associated with the difference in two geometries as discussed in Section~\ref{Polar cap geometry}. Similar shift in the polar cap geometry is seen for the other pole centred at about $\zeta = 160\degr$ in Fig.~\ref{fig:off_lc_30_0_0}(b).

Apart from the expected shift in phase for both the emission regimes between the centred and off-centred case, we notice that the pulse width is also varying. These pulse widths give us an estimate of the size of the polar cap. For instance, in 
Fig.~\ref{fig:off_lc_30_0_0}(a), the width of the 'dip' in the high energy curves decrease as we move from $\zeta = 20\degr$ to $\zeta = 50\degr$. Based on our current understanding of high energy emission, we know that the emission is concentrated in regions closer to the light cylinder (inside or outside) than to the surface of the star because of the spectral sub-exponential cut-off in the GeV range as reported by the Fermi/LAT collaboration \citep{abdo_second_2013}. Opacity in a strong magnetic field as the one present in the polar cap model of \citet{ruderman_theory_1975} would lead to a super-exponential cut-off not seen in gamma-rays. Moreover, recent pulsed TeV photons detection from the Crab pulsar \citep{ansoldi_teraelectronvolt_2016} suggests even another mechanism of radiation, for instance inverse Compton or synchrotron self-Compton from outer parts of the magnetosphere \citep{hirotani_high-energy_2007} or even from the wind \citep{mochol_very_2015}. High-energy emission is minimum at poles which means, that this dip can give us, at least a rough estimate of the differences in the size of the polar caps between the centred and the off-centred case. In part (b) of the same figure we notice the polar cap getting widened (i.e. the high energy emission dip increasing) as we move from $\zeta = 130\degr$ to $\zeta = 160\degr$ for both, the centred and the off-centred case with the width increasing more for the latter as compared to the former. These differences in the widths of the dips of the high energy emission signifies the difference in the size of the polar caps between the two geometries which we highlighted in Section~\ref{Polar cap geometry}.

We also plot the orthogonal case in Fig.~\ref{fig:off_lc_90_90_90} for $\zeta = 80\degr$ to $\zeta = 110\degr$ where $\zeta = 90\degr$ is for the equatorial plane containing both the poles for this inclination. Looking at $\zeta$'s in vicinity of the equatorial plane enables us to look at the emission from both the poles considering one line of sight and hence, we see two peaks of radio emission, one from each pole, with a phase difference of around $180\degr$ between them. Similarly, two high energy emission peaks from two opposite sides of the light cylinder lying orthogonal to the location of the poles are prominent.

\begin{figure}
   \centering
      \begin{tabular}{c}
      \subfloat[ $~\zeta = 20\degr~,30 \degr~,40 \degr~,50 \degr$]{\resizebox{!}{0.6\textwidth}{\input{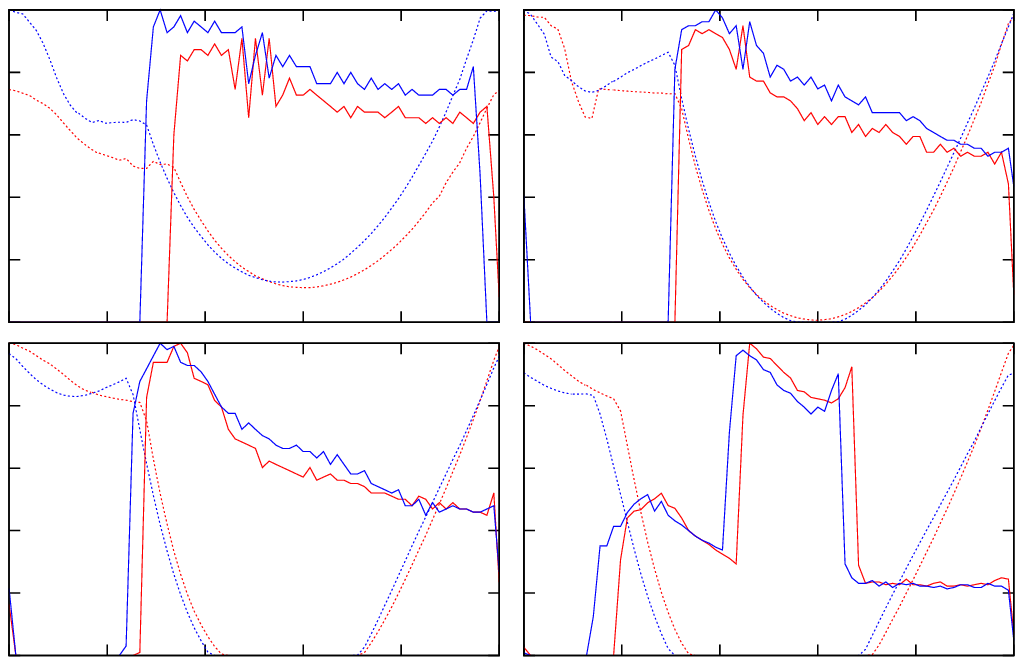}}}\\
      \subfloat[ $~\zeta = 130\degr~,140 \degr~,150 \degr~,160 \degr$]{\resizebox{!}{0.6\textwidth}{\input{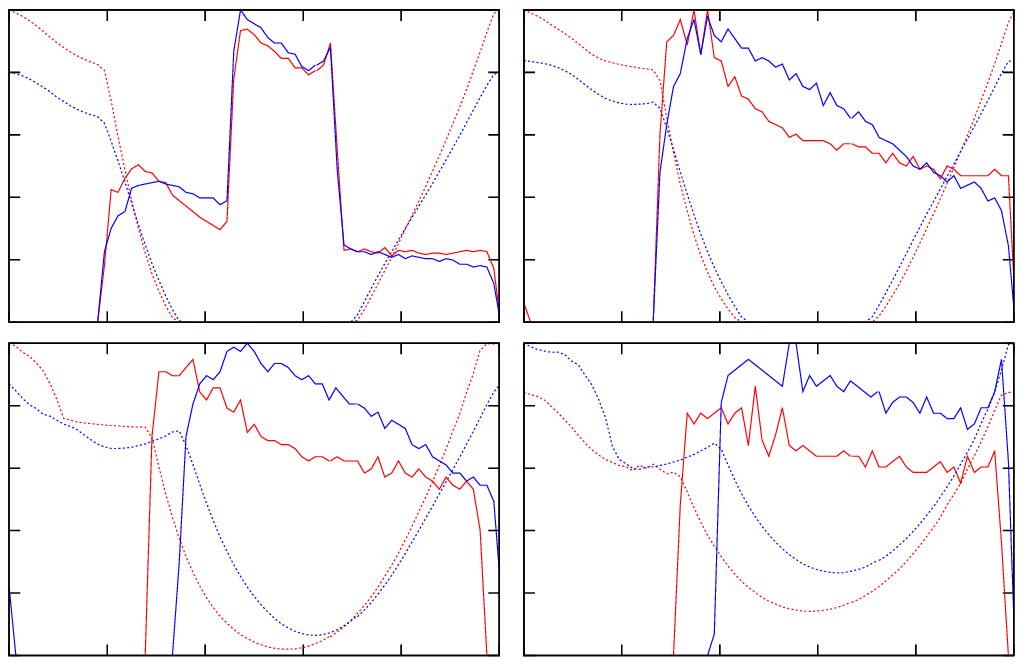}}}    
      \end{tabular}
    \caption{High energy and radio emission light curves with zero phase considered to be the phase of the peak of radio emission for $ (\alpha, \beta, \delta) = (30\degr, 0\degr, 0\degr)$ for centred case ($ \epsilon = 0 $, in red) and off-centred case ($ \epsilon = 0.2 $, in blue) for various line of sights ($\zeta$). In (a) we have $\zeta$ ranging from $20\degr$ to $50\degr$ and (b) has $\zeta$ from $130\degr$ to $160\degr$ in steps of $10\degr$ each. The high energy emission curves are represented in solid lines and the radio emission curves in dashed lines.}
      \label{fig:off_lc_ph0_30_0_0}
\end{figure}

\begin{figure}
   \centering
    \resizebox{!}{0.6\textwidth}{\input{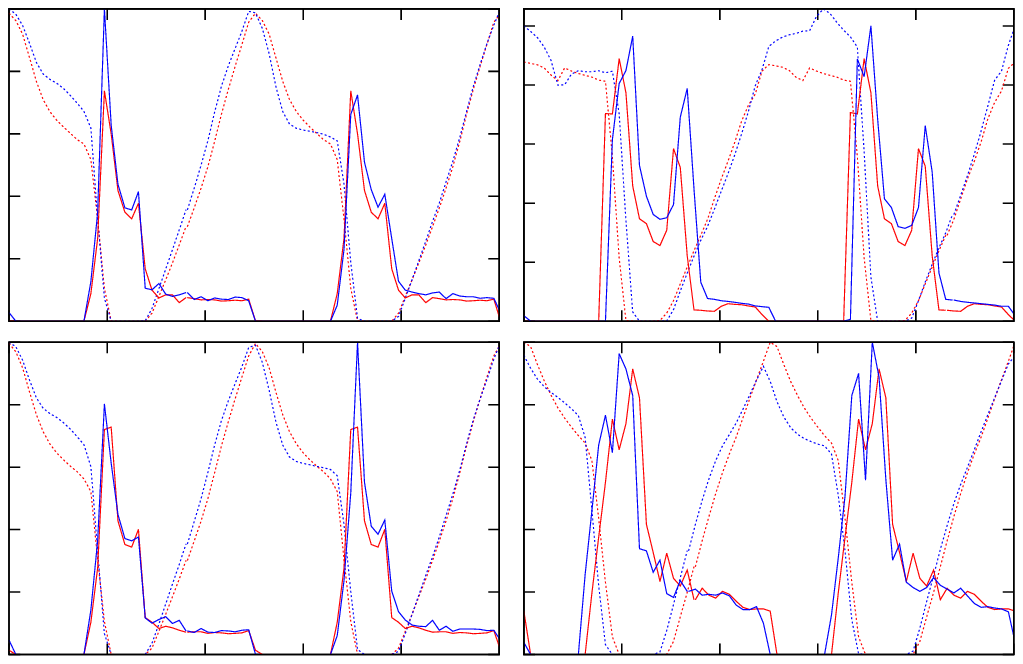}}
    \caption{High energy and radio emission light curves with zero phase considered to be the phase of the peak of radio emission for $ (\alpha, \beta, \delta) = (90\degr, 90\degr, 90\degr)$ for centred case ($ \epsilon = 0 $, in red) and off-centred case ($ \epsilon = 0.2 $, in blue) for various line of sights ($~\zeta = 80\degr~,90 \degr~,100 \degr~,110 \degr$). The high energy emission curves are represented in solid lines and the radio emission curves in dashed lines.}
      \label{fig:off_lc_ph0_90_90_90}
\end{figure}

From an observationalist's point of view, the light curves hold significance only when the emission peak is considered to be the point corresponding to zero phase. It would be convenient to compare the results with observational data if we consider plotting the phase zero light curves. With light curves already available, it is not too difficult to generate phase zero light curves. To make it so, the phase corresponding to each radio emission peak i.e. the location of the maximum intensity of the emission, is put as a reference phase zero and the rest of the curve is shifted accordingly. For the same inclination geometries discussed above, the corresponding phase zero light curves are plotted in Fig.~\ref{fig:off_lc_ph0_30_0_0} and Fig.~\ref{fig:off_lc_ph0_90_90_90} with the colour scheme and other parameters being the same as for above light curves.

In Fig.~\ref{fig:off_lc_ph0_30_0_0} we see the phase lag between the radio emission peak and high energy emission peak varies for both the centred and the off-centred cases as we move to higher $\zeta$'s. It might be of interest to note that in (a) part of the figure, the phase difference between the radio and high energy emission peak decreases with increase in angle of line of sight for both the geometries but comparing them closely reveals that it is faster for the centred case, implying the lag between the radio and high energy emission peaks stays for a longer range of $\zeta$ in the off-centred geometry. While in part (b) of the same figure, the phase difference increases with the angle of line of sight with a faster increase with respect to $\zeta$ in the off-centred geometry.

The phase zero light curves for the orthogonal geometry are shown in Fig.~\ref{fig:off_lc_ph0_90_90_90} which contrast the difference between the centred and off-centred cases clearer than the light curves for the same geometry discussed earlier. We hope to extract more conclusions by comparing the results with observational data at some later stage.

\section{Conclusions}\label{Conclusions}
We extend the radio and high-energy pulsed emission properties of the rotating dipole to an off-centred rotating dipole. The off-centred topology is a reliable approach trying to better fit the broadband spectrum of pulsar radiation and the associated phase-resolved polarisation.

Comparing the polar caps shapes from our approach with the prevalent standard one shows a shift highlighting the difference in size of the polar caps between the two cases and also a phase difference establishing the presence of a time lag between the radio and high energy emission photons. This could explain observational signatures of time lags.

Study of the emission mechanisms with the phase diagrams for the radio and the high energy emission gives strong hints about the site of production of pulsed radiation which could give a better insight in our understanding of the emission mechanism. Also, a comparison of the emission light curves shows phase contrasts between the radio and high-energy profiles for the two geometries. We expect to extract more understanding from the results in future while making a comparison with observational data.

Also, even more constraints could be gained for the emission mechanisms by inspection of the polarization features of an off-centred dipole \citep{petri_polarized_2017-1}. Our next goal would be a detailed and accurate analysis of phase-resolved polarization using this approach in vacuum. We plan also to extend our study to pulsar force-free magnetosphere.

\section*{Acknowledgements}
We are grateful to the referee for their comments and suggestions. This work has been supported by the French National Research Agency (ANR) through the grant No. ANR-13-JS05-0003-01 (project EMPERE) and by the University of Strasbourg IdEx programme.


\appendix

\section{Transformation of coordinates}\label{transformation_coordinates}
\subsection{From magnetic axis frame to rotation axis frame :}
To ensure that no points are lost attributed to the complexity of the off-centred geometry, we begin all our calculations considering the initial coordinates with respect to the magnetic axis frame and then, make a transformation to the rotation axis frame for further simulations. The transformation between the two frames requires two steps of rotation; first along the $y$-axis and then second one along the $z$-axis. Rotation matrix for such transformations \citep{arfken_mathematical_2013}, respectively, are: 

\begin{math}
\mathbf{R}(y) = \begin{bmatrix}
 \cos{\theta} & 0 & \sin{\theta} \\
 0 & 1 & 0 \\
 -\sin{\theta} & 0 & \cos{\theta}
\end{bmatrix}
\end{math}

and 

\begin{math}
\mathbf{R}(z) = \begin{bmatrix}
 \cos{\phi} & -\sin{\phi} & 0 \\
 \sin{\phi} & \cos{\phi} & 0 \\
 0 & 0 & 1
\end{bmatrix}
\end{math}

where $\theta$ and $\phi$ are the polar and the azimuth angles for poles with respect to the rotation axis. We get the coordinates in rotation axis frame ($\mathbf{x}$) using above transformation matrices and the coordinates in the magnetic axis frame ($\mathbf{x'}$) using $\mathbf{x} = \mathbf{R}(z) \mathbf{R}(y) \mathbf{x'}$.

\subsection{Between Cartesian coordinate system and spherical coordinate system :}\label{app:cc_sc}
To transform the Cartesian coordinates ($x, y, z$) to spherical coordinate system ($r, \theta, \phi$), we use the standard transformation \citep{griffiths_introduction_2013}:\\
\begin{math}
 r = \sqrt{x^2 + y^2 + z^2}\\
 \theta = \cos^{-1}{\frac{z}{r}}\\
 \phi = \tan^{-1}{\frac{y}{x}}\\
\end{math}
and the following for the inverse transformation :\\
\begin{math}
  x = r \sin{\theta} \cos{\phi}\\
  y = r \sin{\theta} \sin{\phi}\\
  z = r \cos{\theta}\\
\end{math}
with $\theta\in [0,\pi]$, $\phi\in [0,2\pi)$ and $r\in[0,\infty)$.

\section{Transformation of magnetic field from spherical coordinate system to Cartesian coordinate system }\label{transformation_magnetic_field}
To transform the magnetic field from the spherical coordinates ($B_{r}, B_{\theta}, B_{\phi}$) to Cartesian coordinates ($B_{x}, B_{y}, B_{z}$), following transformation scheme was used :\\
Writing the total magnetic field $\mathbf{B}$ in terms of spherical coordinates using spherical unit basis vector ($ \mathbf{\hat{r}}, \boldsymbol{\hat{\theta}}, \boldsymbol{\hat{\phi}} $),
\begin{equation}\label{mf_sc}
  \mathbf{B}(r, \theta, \phi) = B_{r}(r, \theta, \phi) \mathbf{\hat{r}} + B_{\theta}(r, \theta, \phi) \boldsymbol{\hat{\theta}} + B_{\phi}(r, \theta, \phi) \boldsymbol{\hat{\phi}}.
\end{equation}
Now, to convert the spherical unit basis vector to Cartesian unit basis vector, the transformation equations \citep{griffiths_introduction_2013} are :\\
\begin{math}
  \mathbf{\hat{r}} = \sin{\theta} \cos{\phi} \mathbf{\hat{x}} + \sin{\theta} \sin{\phi} \mathbf{\hat{y}} + \cos{\theta} \mathbf{\hat{z}}\\
  \boldsymbol{\hat{\theta}} = \cos{\theta} \cos{\phi} \mathbf{\hat{x}} + \cos{\theta} \sin{\phi} \mathbf{\hat{y}} - \sin{\theta} \mathbf{\hat{z}}\\
  \boldsymbol{\hat{\phi}} = - \sin{\phi} \mathbf{\hat{x}} + \cos{\phi} \mathbf{\hat{y}}.\\
\end{math}
Using above transformation equations in Eqn.~(\ref{mf_sc}) and then, comparing all individual components of the magnetic field with,
\begin{equation*}
 \mathbf{B}(r, \theta, \phi) = B_{x}(r, \theta, \phi) \mathbf{\hat{x}} + B_{y}(r, \theta, \phi) \mathbf{\hat{y}} + B_{z}(r, \theta, \phi) \mathbf{\hat{z}}
\end{equation*}
we get, the individual components of the magnetic field in Cartesian coordinate system as, $B'=T\,B$ where $B'= (B_{x}, B_{y}, B_{z})$ and $B = (B_{r}, B_{\theta}, B_{\phi})$ are magnetic fields in Cartesian and spherical coordinates respectively, and $T$ is the transformation matrix written as :\\
\begin{math}
\begin{bmatrix}
 \sin{\theta} \cos{\phi} & \cos{\theta} \cos{\phi} & -\sin{\phi} \\
 \sin{\theta} \sin{\phi} & \cos{\theta} \sin{\phi} & \cos{\phi} \\
 \cos{\theta} & -\sin{\theta} & 0
\end{bmatrix}
\end{math}

\bsp	
\label{lastpage}
\end{document}